\documentclass{article}

\pdfoutput=1

\usepackage[T1]{fontenc}
\usepackage{times}

\usepackage{fancyvrb}
\CustomVerbatimEnvironment{CodeVerbatim}{Verbatim}{numbers=left, samepage=false}
\usepackage{float}

\usepackage{amsmath}
\usepackage{amscd}
\usepackage{color}
\usepackage{multicol}
\usepackage[margin=10pt,
        font=footnotesize,
        labelfont=bf,
        tableposition=top]{caption}
\usepackage{ifthen}
\usepackage{listings}
\usepackage{sidenotes}
\usepackage[utf8]{inputenc}

\usepackage{bookmark}
\usepackage[superscript]{cite}
\usepackage{texlinks}

\usepackage{xcolor}
\usepackage{xspace}

\usepackage{Sweave}

\DeclareRobustCommand{\DJ}{\marginnote{\large{\textcolor{magenta}{\sf Derek}}}}

\DeclareRobustCommand{\WN}{\marginnote{\large{\textcolor{blue}{\sf Bill}}}}

\definecolor{hellgelb}{rgb}{1,1,0.9}
\definecolor{colKeys}{rgb}{0,0,1}
\definecolor{colIdentifier}{rgb}{0,0,0}
\definecolor{colComments}{rgb}{1,0,0}
\definecolor{colString}{rgb}{0,0.5,0}

\usepackage{hyperref}

\begin{document}

\title{The CESAW dataset: a conversation}
\author{Derek M. Jones\\Knowledge Software\\derek@knosof.co.uk
\and William R. Nichols\\Carnegie Mellon University\\wrn@andrew.cmu.edu}
\maketitle

\begin{abstract}
An analysis of the 61,817 tasks performed by developers working on 45
projects, implemented using Team Software Process, is documented via
a conversation between a data analyst and the person who collected,
compiled, and originally analyzed the data.  Five projects were
safety critical, containing a total of 28,899 tasks.

Projects were broken down using a Work Breakdown Structure to create
a hierarchical organization, with tasks at the leaf nodes.  The WBS
information enables task organization within a project to be
investigated, e.g., how related tasks are sequenced together.

Task data includes: kind of task, anonymous developer id, start/end
time/date, as well as interruption and break times; a total of
203,621 time facts.

Task effort estimation accuracy was found to be influenced by factors
such as the person making the estimate, the project involved, and the
propensity to use round numbers.
\end{abstract}

\section{Introduction}

This paper takes the form of a conversation between William Nichols
who was the technical lead member of the team who obtained and
analysed, the data, and Derek Jones who reanalyzed the data.

Data analysis is an iterative process; ideas may have been suggested
by discussions with those involved before the data arrives, and new
ideas are suggested by feedback from the ongoing analysis.  Most
ideas go nowhere; failure of the data to support an idea is the norm.
Analysts who are not failing on a regular basis never discover
anything.

The reason for analyzing data is to obtain information that is useful
to those involved with the processes and systems that produced the data.

Any collection of measurement data contains patterns, and some of
these may be detected by the statistical techniques used.  Connecting
patterns found by data analysis, to processes operating in the world
requires understanding something about the environment and practices
that generated the data.

If the person doing the data analysis is not intimately familiar with
the environment and practices that generated the data, they either
have to limit themselves to generalities, or work as a part of a team
that includes people who have this knowledge.

As the conversation progressed, the narratives created as possible
explanations for the patterns found in the data evolved; readers are
presented with a semi-structured story fitted together after the
event.

One purpose of this conversation is to provide an introduction to the
CESAW data.

The data is available for download at:

\url{https://kilthub.cmu.edu/articles/CESAW_Project_Data_Fact_sheets/9922697}

\subsection{Stumbling onto data}

\DJ

Detailed software project data is very difficult to acquire, and I
was surprised and delighted to discover that a report by Nichols,
McHale, Sweeney, Snavely and Volkman \cite{Nichols_18} included a
release of a very large collection of project data.  This data
contained a lot more information than was discussed in the report,
and Bill Nichols was very prompt in answering the questions I asked
about the data, and agreed to work on publishing this conversation.

Table~\ref{CESAW-files:tab} shows the number of rows and columns
contained in the largest files in the CESAW dataset.

\begin{table}
\begin{center}
\begin{tabular}{lrr}
\bfseries File & \bfseries Rows & \bfseries Columns \\[0.5ex]
\itshape CESAW\_defect\_facts.csv & 35,367 & 14\\
\itshape CESAW\_size\_facts.csv & 15,942 & 16\\
\itshape CESAW\_task\_fact.csv & 61,817 & 12\\
\itshape CESAW\_time\_fact.csv & 203,621 & 15\\
\end{tabular}
\end{center}
\caption{Number of rows and columns in CESAW's largest files.
{\itshape CESAW\_defect\_facts.csv}: All defects found during development.
{\itshape CESAW\_size\_fact\_sheet.csv}: All size information from the projects' work breakdown structures.
{\itshape CESAW\_task\_facts.csv}: All task information connected to the projects' work breakdown structures.
{\itshape CESAW\_time\_facts.csv}: Time log entries from projects connected to work breakdown structure.
}
\label{CESAW-files:tab}
\end{table}

My professional background is compiler writing and source code
analysis.  Over the last ten years I have collected and analyzed 600+
software engineering datasets and made them publicly available
\cite{Jones_20}.

\WN

Although my background in Physics, I've been doing software
development and software engineering my entire professional life.
Since 2001 I've been tracking individual effort on project work using
the Team Software Process framework and since 2006 I've been doing
this with the Software Engineering Institute.  The CESAW dataset is a
small subset of my project inventory that was selected because the
developers used some specific software assurance tools. The work is
tracked in sufficient detail that I was able to isolate the direct
effort to use those tools, the defects found by those tools, and the
effort required to fix those defects. In short, the data is designed
for both work tracking and process improvement. I was able to show
that using source code analysis tools don't slow down initial
development because the find and fix rates are so much faster than
dealing with the issues that would likely have been found in test.
The challenge isn't the math in the analytics, it is becoming
familiar enough with the data to use it. 

\subsection{The conversation}

\DJ

Getting the most out of data analysis requires domain knowledge.
Bill has that knowledge, but is a busy man.  I find the best way to
get a busy person to talk to me, is to tell them things about the
data that they find interesting and useful.

My top priority is to find something in the data that the domain
expert finds interesting.  The boring, but necessary, stuff can be
done later.  This approach is one reason for the disjoint
organization of the paper.

\section{SEMPR and the CESAW subset}

\begin{figure}
\begin{center}
\includegraphics{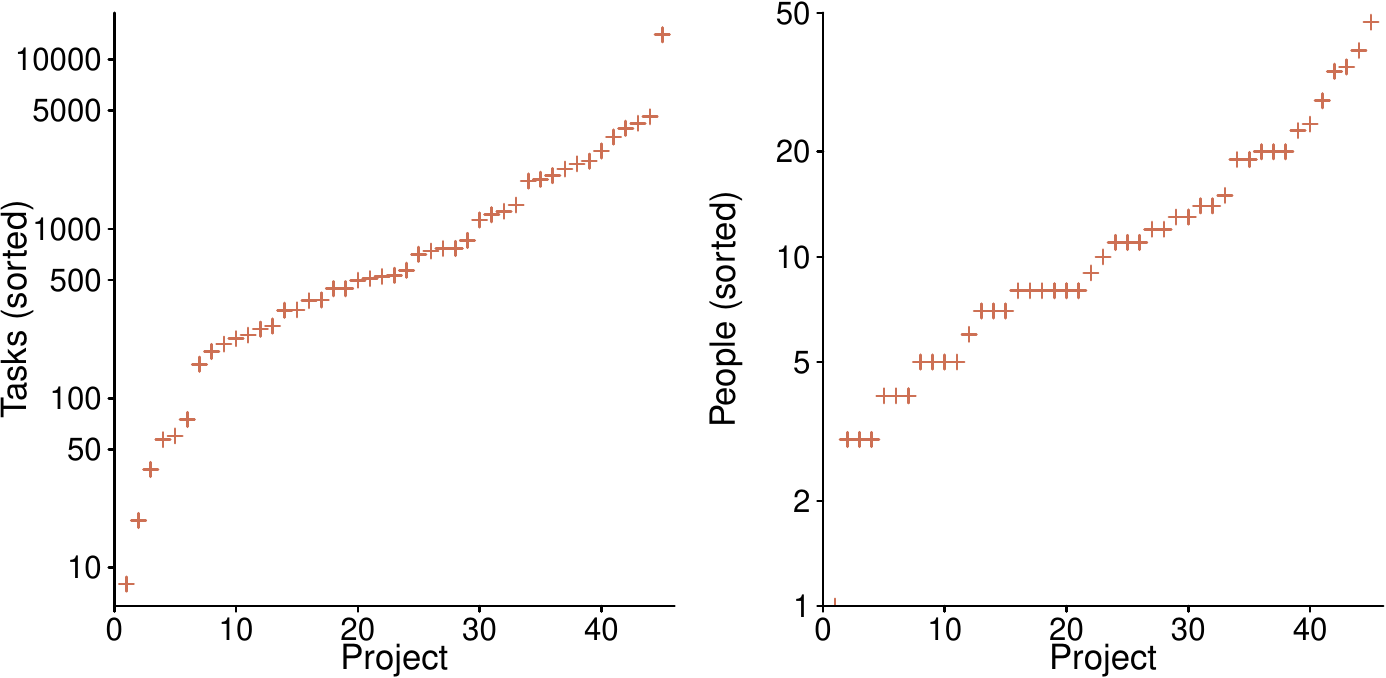}
\end{center}
\caption{Left: Number of tasks (in sorted order) recorded for each project; right: Number of people (in sorted order) who worked on each project.}
\label{proj-task:fig}
\end{figure}

\WN

The SEMPR data contains about 1,000 software project cycles from 85
companies. SEMPR (the Software Engineering Measured Process
Repository) is a data warehouse populated from Process Dashboard
\cite{PD-Tuma} project data files.  CESAW (Composing Effective
Software Assurance Workflows) was a research project that examined
the cost/benefits of static analysis.  The CESAW report included
three companies from the SEMPR data for which we could identify the
use of static analysis tools: five projects from a company working in
Avionics (safety critical), 16 from a company working in business
intelligence,and 15 from a company working in design automation.
These companies use Team Software Process \cite{TSP-2000} (TSP) for
software development, and the Process Dashboard for data collection.

\DJ

Figure~\ref{proj-task:fig} shows the number of tasks contained in
each of the projects, along with the number of people involved in
each project.  Project implementation occurred between 2010 and 2017,
see figure~\ref{elapsed-time:fig}, right plot, for project durations.

The analysis in this paper involves the task related data contained in the files:
{\itshape CESAW\_task\_facts.csv} and
{\itshape CESAW\_time\_facts.csv}.
The data in the code related files:
{\itshape CESAW\_defect\_facts.csv} and
{\itshape CESAW\_size\_fact\_sheet.csv} are not analyzed.

\subsection{Work Breakdown Structure}

\DJ

All projects used the Work Breakdown Structure (WBS) to subdivide a
project into ever smaller components.  The smallest component, a WBS
element, contains a sequence of tasks to be performed to implement
that component.  These components are denoted by integer values in
the column \texttt{wbs\_element\_key}.

\begin{figure}
\begin{center}
\includegraphics{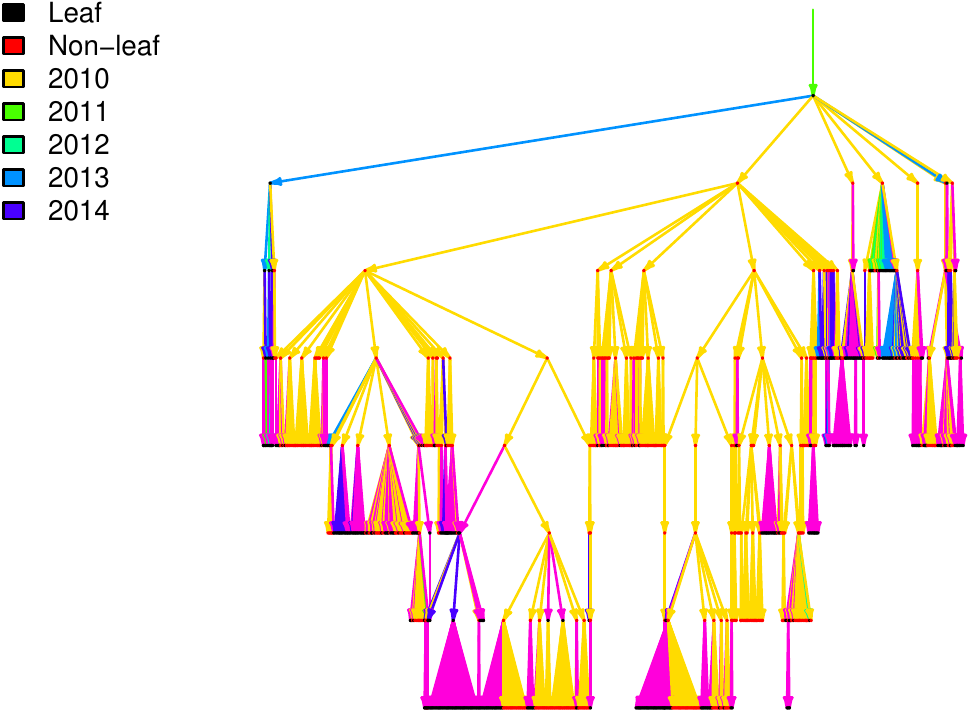}
\end{center}
\caption{The Work Breakdown Structure tree of the 961 wbs elements in project 615; leaf WBS elements (in black, contain a sequence of tasks), non-leaf WBS elements reference other WBS elements (in red; or terminate with no corresponding tasks), other colored edges denote year in which associated leaf tasks were first started.}
\label{WBS-tree:fig}
\end{figure}

A Work Breakdown Structure is a hierarchical tree;
figure~\ref{WBS-tree:fig} shows the WBS tree for project 615.  A WBS
is intended to support the evolution of a system through incremental
development, with the requirement that the system be broken down into
a structure that shows which capabilities will be satisfied by a
specific increment.  MIL-STD-881E \cite{MIL-STD-881E_20} is the US
Department of Defense standard for Work Breakdown Structures.

In practice there are three major perspectives on the approach to the
use of a work breakdown structure:

Deliverable-oriented: The approach defined by the PMI body of
knowledge \cite{PMBOK_13}, in which decomposition is structured by
the physical or functional components of the project.  The major
project deliverables are the first level of the WBS.

Activity-oriented: This focuses on the processes and activities in
the software project.  The major life cycle phases are the first
level of the WBS.

Organization-oriented: This focuses, like the activity-oriented
approach, on the project activities, but groups them according to
project organizational structure. The subprojects or components of
the project are the first level of the WBS.  Subprojects can be
identified according to aspects of project organization as created
subsystems, geographic locations, involved departments or business
units, etc.  An organization-oriented WBS may be used the context of
distributed development.

\begin{figure}
\begin{center}
\includegraphics{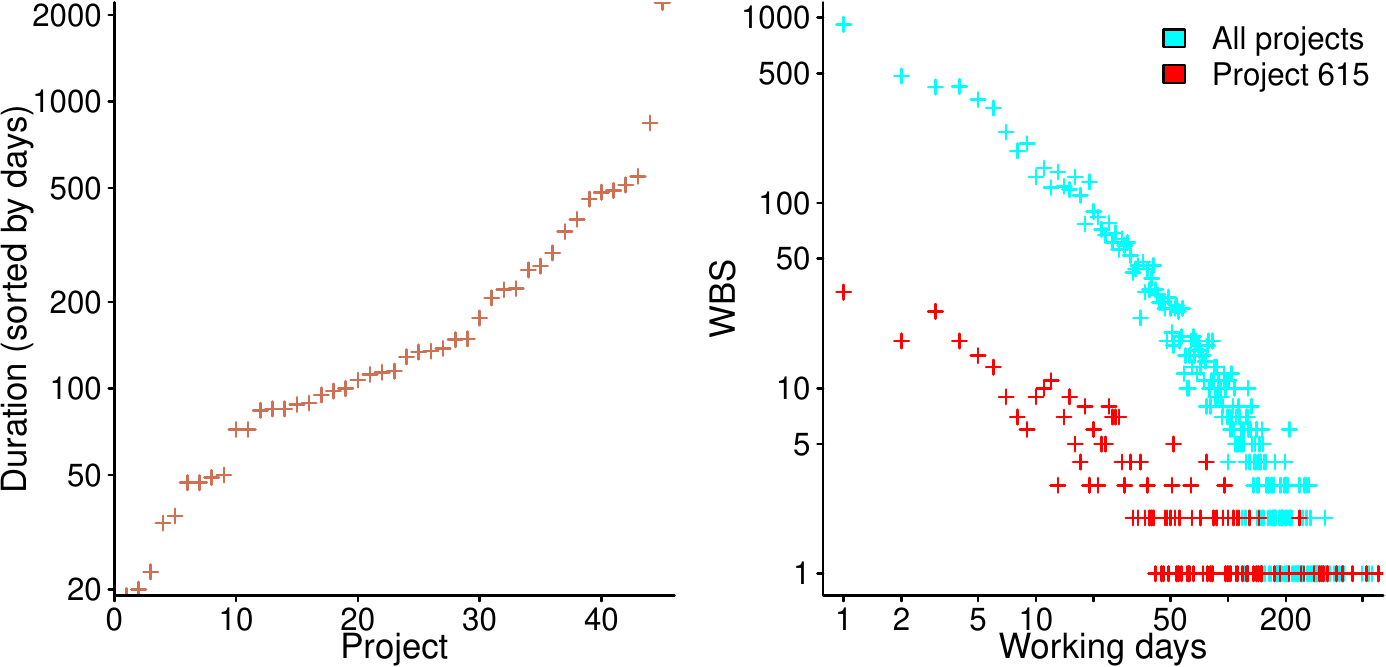}
\end{center}
\caption{Left: Duration of each project, sorted by days; right: Number of WBS whose implemented occurs over a given number of working days, for all projects and project 615.}
\label{proj_WBS-d:fig}
\end{figure}

Figure~\ref{proj_WBS-d:fig}, left plot, shows the duration of each
project; the right plot shows the number of WBS whose implementation
took a given number of elapsed working days, for all projects and
project 615.

\WN

The WBS should structure the overall work, and is a basis for
tracking progress to completion. In our approach, the top levels can
be product based (modules) or functionally based (requirements). It
doesn't matter. The leaf elements represent the process elements of
the work. There will always be some grey areas on where to place
integration, test, and deployment. These are often process leafs off
a higher level node.  In the Dashboard, task is related to the WBS
element through the Plan item.  For the fact sheets I flattened the
relationship, but the warehouse has more indirect links. The
warehouse was optimized for adding data to a dynamic project. That
sometimes makes extracting the data more work.

Tasks in the WBS are built from process workflow definitions.  The
phase list has a natural sequence of work order, requirements before
design, before code, before test. 

The product is physically decomposed (sometimes modules, more often
features or changes that cut across physical modules). The final
level of decomposition is task by phased activity. This is done
automatically. The work package (WBS element) is sized, and the
effort is distributed across the phases using historical phase effort
ratios. 

\DJ

Why are some leaf WBS elements not associated with tasks (in
figure~\ref{WBS-tree:fig} these are the red nodes that terminate
without an outgoing child link)?

\WN

First, when I built the task sheet, I only included tasks that that
had time logged. I took the effort by aggregating the time log
entries by task then got the other task data from task tables. So no
tasks only tells us that these tasks were not worked.  Next, I went
to the context information to look into this. Unfortunately, I cannot
share that material because it cannot be anonymized. (there should be
some future work joining that context data) 

I see two distinct types of WBS component that these could be.

1) placeholders that were never estimated,

2) WBS components that were estimated but never started and completed.

Type 1) seem to be work that is kind of speculative, for example a
placeholder for fixing issues returned from test that someone else
(not in the plan) has performed. Another example included fixing
security issues or defects. This is often speculative rework. The
tasks did not look like software components, but placeholders to make
the plan more realistic by reminding them , "hey, these were things
did in the past, but they might not be needed, but we better put them
in, so we don't forget about them". 

Type 2) can be different.  If you look at the task sheet, some of the
tasks have an end date, but no start. These were closed off without
working them. If the task was not worked and never closed, it never
made the task sheet.  I can only make an educated guess that these
were either not needed, de-prioritized and not worked, or they ran
out of time and didn't get to these before the plan was submitted.
That is, cancelled tasks. There is no direct evidence for why these
things happened. The data only tells us something odd happened. There
may be explanations in a team meeting report. 

\DJ

\begin{figure}
\begin{center}
\includegraphics{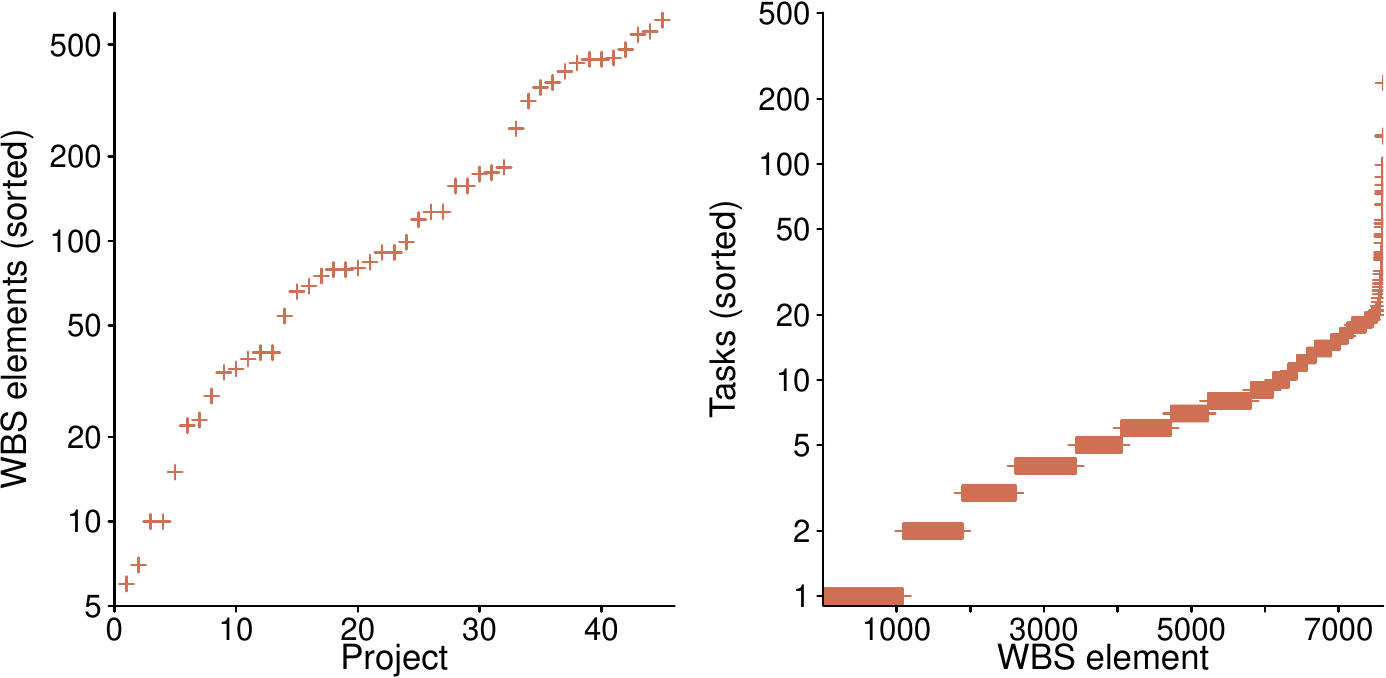}
\end{center}
\caption{Left: Number of WBS (in sorted order) contained in each project; right: Number of tasks (in sorted order) contained in each WBS for all projects.}
\label{proj-WBS:fig}
\end{figure}

Figure~\ref{proj-WBS:fig}, left plot, shows the number of leaf WBS
elements in each project (in sorted order), and the right plot shows
the number of tasks in each WBS element (for all projects in both
cases).

While most WBS involve a few tasks (across all projects),
in some projects many WBS contain many tasks; see
figure~\ref{wbs-people:fig}, right plot.

Derek: wbs 80208 has 7,143 tasks, and wbs 1 has 3,796 tasks.  Might
these be database extraction mistakes?

\WN 

I won't claim there are no mistakes in extraction, but these are
heterogeneous projects. Some are very short while several that are
multi-year efforts. I don't see any reason to believe a priori that
the structures should be all that similar. 

I want to look at the structure one more time, there may be some
multiple counting because of the structure. 

The \textsf{Misc} phase is used as a catchall for tasks that pop up
or don't fit into a standard workflow category.  \textsf{Misc} often
includes orphaned entries. Req Inspection also depends upon a
specific team raises some questions that would require context.
Nonetheless, an inspection should normally be part of a sequence of
tasks, i.e. a workflow.

As you have recognized, there is a lot of opportunity to audit the
data and inconsistencies sometimes appear.  It is entirely possible
there is a bug in my extraction software, but I can always go to the
original source to verify.

Multiple people may be on a plan item (for example, a code
inspection) each would have a task.  Note, WBS\_Element may not
always be a physical product. Integration or System test may stand
alone without a direct mapping to the design and code products.  The
integration content must be inferred from timestamps.

The leaf element usually has a process associated with. WBS\_item(X)
Process\_Vector == task list.  But sometimes they needed to add a
task with only a single process phase.  To verify I'll have to look
more carefully at that data.

This planned item was worked in two sessions, one of 81 minutes, the
(other the next day) of 131 minutes.  That's a little over an hour
and a little over two hours.  Those are very reasonable work times.
Neither work session recorded interrupt time, but for 1-2 hour
sessions, that is also reasonable.

It is pretty normal that a task is worked on in short bursts over
multiple days and this is a reason that date stamps are not very
useful for measuring effort on small tasks. I've been trying to get
the Agile community (using Rally, Version One, or other tools) to
recognize this, to no avail.

\subsection{Team Software Process}

\WN

Team Software Process breaks down the creation of software into
phases defined by a primary activity. This phase structure does not
imply a waterfall project structure, instead it reflects the natural
progression of software through a sequence of activities. After a
piece of software completed a phase, any escaping defects must be
recorded.  The phase, as an accounting mechanism, thus helps to
isolate rework.  Table~\ref{TSP-phases:tab} shows some typical phases
in a TSP project.

\begin{table}
\begin{center}
\begin{tabular}{ll}
\bfseries Description& \bfseries Phase Type \\[0.5ex]
Detailed-level design& Creation \\
Personal review of the detailed design& Appraisal \\
Unit test case development& Creation \\
Peer inspection of the detailed design& Appraisal \\
Writing the source code& Creation \\
Personal review of the source code& Appraisal \\
Peer inspection of the source code& Appraisal \\
Developer unit test execution& Failure \\
Integration test& Failure \\
System test& Failure \\
User acceptance test& Failure \\
Product life& Failure \\
\end{tabular}
\end{center}
\caption{Typical phases of a TSP project, and the types of actions that can occur that are applicable to the project team.}
\label{TSP-phases:tab}
\end{table}

TSP is a superset of PSP phases. PSP focuses on the individual tasks
a developer performs to write a small piece of code. PSP has little
with requirements, but understanding the assignment, which is roughly
similar to a small user story. PSP also doesn't have major
integration or regression/system test. 

The teams define their own workflows and are free to define task
categories, but most are at least loosely based on ISO
12207\cite{ISO_12207} lifecycle phase definitions. A colleague has
suggested we describe these as ISO 15504\cite{ISO_15504} lifecycle
activities to avoid confusion with "waterfall".  Think of the
activities as the steps that single piece of software must proceed
through from an idea to running code. A DevOps pipeline helps to
visualize this.  Each story goes some set of activities such as
requirements analysis, design, code, various reviews or inspections,
unit test, integration, regression test, and deployment.  The value
of TSP is that it captures and places into context all the
development work within a project team.

\subsection{How SEMPR tracked work}

\WN

The Process Dashboard \cite{PD-Tuma} was the data collection tool and
developers used this tool as a work log. 
But the dashboard served several purposes.  
It was used to plan the project, 
to gather data during project execution, 
to monitor project performance, and 
to analyze the project during cycle or project post-mortems.
The planning included 
building a Work Breakdown Structure, 
estimate WBS elements by size or effort,
define  work process flows and processes, 
apply a workflow to convert some WBS elements into a tasklist, 
record project staff and estimate effort available per week,
distribute tasks to individuals,  and
build a schedule. 

Each task was assigned to an individual. The teams then sequenced
upcoming work to manage important coordination events and balance the
workloads.  The resulting straw plan demonstrated that work was at
least possible.

During execution, individuals work the task at the top of their list.
They used an electronic stopwatch to record time, and marked the task
complete when done. Any defects discovered after "done" were recorded
as defects.  As work deviated from the plan, tasks would be reordered
or reassigned to keep the plan in balance. 

At regular intervals, sometimes weekly, but at least at releases, the
Process Dashboard data was exported, shipped to SEI, then imported
into a Process Dashboard data warehouse \cite{PDDW-Tuma} (SEMPR).  We
extracted the data into fact sheets using R scripts and SQL code.  

\subsection{Data collected}

\DJ

Both the Nichols' report \cite{Nichols_18}, and the README included
with the data, provide a basic overview of the information contained
in the columns of each file.  Table~\ref{CESAW-columns:tab} lists
some of the columns and information contained in the files, while
table~\ref{CESAW-files:tab} list the number of rows and columns in
the largest files.

\begin{table}
\begin{center}
\begin{tabular}{ll}
\bfseries Column & \bfseries Information\\[0.5ex]
\texttt{person\_key} & unique developer ID\\
\texttt{phase\_key} & unique ID for the process phase in which the task work was done\\
\texttt{phase\_short\_name} & readable name for process workflow phase\\
\texttt{plan\_item\_key} & unique ID for the planned item associated with the task\\
\texttt{process\_name} & readable name for process workflow\\
\texttt{project\_key} & unique project ID\\
\texttt{task\_actual\_start\_date} & time stamp for first work on task\\
\texttt{task\_actual\_complete\_date} & time stamp when task was completed\\
\texttt{task\_actual\_time\_minutes} & total effort actually logged for the task\\
\texttt{task\_plan\_time\_minutes} & total effort planned for the task\\
\texttt{team\_key} & unique team ID\\
\texttt{wbs\_element\_key} & unique work breakdown structure element ID\\
\end{tabular}
\end{center}
\caption{Some of the columns in the data files, and the information they contain.}
\label{CESAW-columns:tab}
\end{table}

To make full use of the information collected it needs to be mapped
to developer work practices.  Inside knowledge is needed.

\WN

Teams collected an enormous amount of data.  We had a list of every
task performed, the effort spent, where defects were injected, where
they were found, how long they took to fix, and so forth.  That's a
lot of data, but it doesn't tell you the work domain,
the product,
the trade-offs between time and functionality, 
the definition of done,
how the process steps were executed,
the tools used, 
the programming languages,
why the plan was changed,
the basis for estimates,
and so forth.
This was not a problem for the teams because they knew all this. But
those of us trying to use the data have some big information gaps. We
also have other project artifacts including reports, meeting minutes,
and post-mortem analysis.  But these were not automatically stored in
the SEMPR.  We cannot release those because they include a lot of
identifying information. 

\DJ

The information in the files
\textit{CESAW\_time\_fact.csv} and
\textit{CESAW\_defect\_facts.csv} corresponds to PSP's time recording
and defect recording log, respectively. The
\textit{CESAW\_task\_fact.csv} file aggregates the time fact
information for each task, e.g., a sum of the time deltas and range
of start/end dates. 

\WN

The CESAW date, like everything in SEMPR, uses the TSP framework.
Some teams use the standard TSP data framework, which includes
traditional PSP as a subset, but there were some variations. One team
that had systems engineering responsibilities elaborated on the
requirements, including personal review of requirements and high
level (architecture and structural) design.  Isolating Static
Analysis was a little tricky because no one defined it as a separate
phase.  Fortunately, the defect descriptions helped to identify the
source of defects, so I was able to use the defect find and fix time.
Other static analysis work turned up in Integration Test and System
Test.  It depended on the type of tools being used and what type of
issues it was designed to find.

\DJ

The report lists three companies (A, B, and C), but there is no
mention of the D company that appears in the data.

\WN

Company D was removed from our study because we could not reliably
isolate the use of static analysis tools.

\section{Initial analysis}

\DJ

The CESAW dataset shares some similarities to a company dataset
previously analysed by Derek, the SiP dataset \cite{Jones_19a}, and
the initial analysis investigated the same relationships.

As the analysis progressed the structures present in the data began
to be discovered, suggesting possible new patterns of behavior (some
of which were found to be present).

To provide some focus for the initial analysis project 615 was chosen
as a test bed because it was the project containing the most tasks; this
project contained safety critical components.

\WN

Project 615 is an end to end development so there is a lot of
requirements engineering work. Many projects do primarily code.  That
is, most teams are given requirements or stories and work mostly in
the implementation phase. In this project we get to see more of the
life cycle. 

\subsection{A first regression model}

\DJ

Before starting work on a task, the person involved makes an estimate
of the expected duration (in minutes); the actual time taken is
recorded on completion.
What is the relationship between estimated (the
\texttt{task\_plan\_time\_minutes} column in the data) and actual
time (the \texttt{task\_actual\_time\_minutes} column)?

\begin{figure}
\begin{center}
\includegraphics{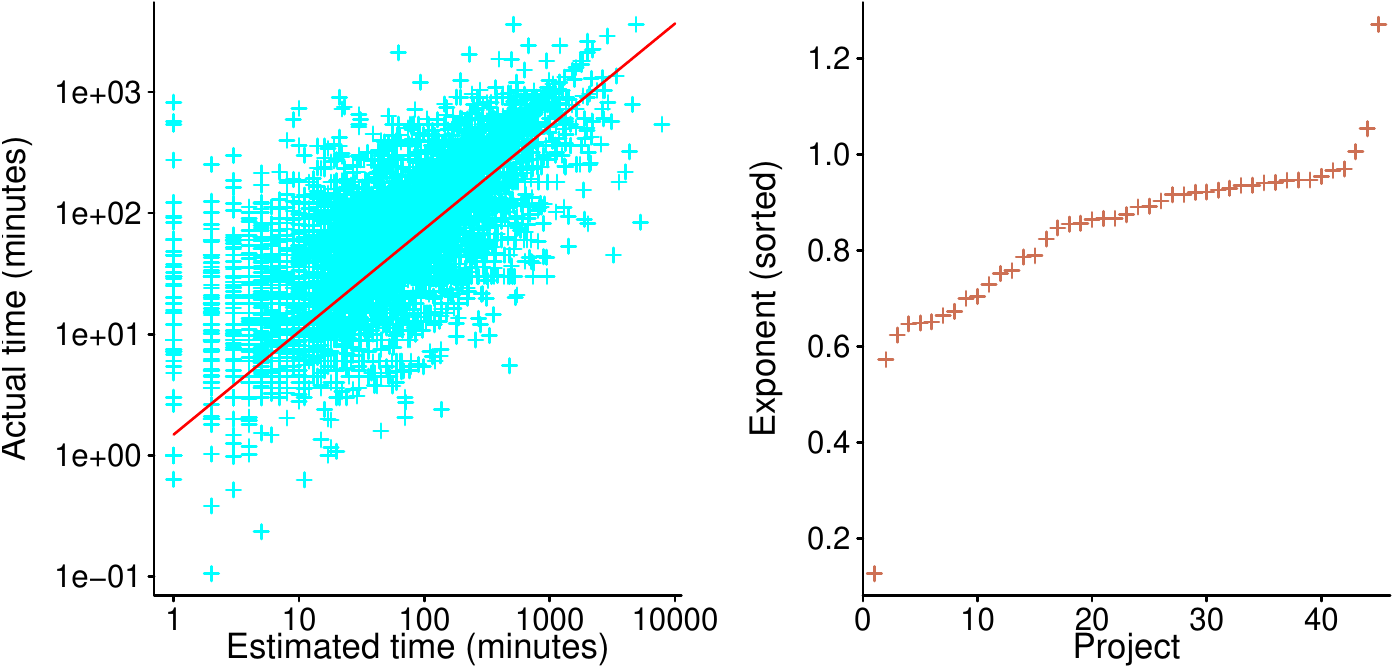}
\end{center}
\caption{Left: Estimated task work time (in minutes) against actual task work time for the 61,817 tasks in all projects, with fitted regression model of the form $Actual \propto Estimated^{0.85}$; right: Exponent of models fitted (in sorted order) to every project.}
\label{est_act:fig}
\end{figure}

Analysis of estimates made for other projects has found that a
power-law provides a good fit.  Figure~\ref{est_act:fig}, left plot,
shows estimated time against actual time to complete a task, for all
61,817 tasks, with a fitted regression model (in red); the fitted
equation has the form:

\begin{equation}\mathit{Actual\_mins}= 1.5 \mathit{Estimated\_mins}^{0.85}\label{first_regression}\end{equation}

This model explains 52\% of the variance present in the data.  The
first equation fitted to the SiP data has the form:
$\mathit{HoursActual}=1.1\mathit{HoursEstimate}^{0.87}$, and
explained 70\% of the variance.

\WN

Keep in mind that this is direct effort, not wall clock or calendar
time. We coached the teams to aim for consistent estimation. As long
as they had a consistent distribution, accuracy could be corrected
for with regression and the high and low estimates would balance out. 

\DJ

How do individual projects compare against the aggregate model?
Figure~\ref{est_act:fig}, right plot, shows the fitted exponent value
(in equation~\ref{first_regression}) for each project.

This regression model can be extended to include other columns in the
data, e.g., organization, project, individuals, and task phase.
However, there are larger project structural factors that probably
ought to be investigated first.  For instance, the TSP phases occur
in a specified order, and the relationships between tasks performed
in a sequence may need to be taken into account in a realistic model.

\WN

I've puzzled for a while how to use the data for observational
studies of programmer/developer performance.  Absent an objective
measure of task size, how could we infer productivity effects for
long or short work sessions, or having multiple tasks open?  I've
tried to set up structural equation models, SEM, but the structure
doesn't seem quite suit the models and data.

Hypothetically, the optimum number of open tasks should be 1, and the
ideal work session without breaks would be longer than 30 minutes,
but shorter than 3 hours.  There should be a start up penalty for
performing work after a long break (say several hours). 

Because this is observational, these might not be available from the
data, but I can imagine starting with the estimate, then calculating
the effect of the adjustment factors. 

\DJ

What is productivity?  When repeatedly creating the same item, such
as a widget, the number of widgets created per hour is an obvious
measure of productivity.  Each task on a software development project
is different, yes tasks may have similarities, which means the effort
involved is different for each of them.  Monitoring developer brain
activity is another possibility, but without a model of how the brain
solves problems we can do little more than say somebody is thinking.

While one task is ideal, in practice there will be road-blocks.
So it might be useful to be able to take on multiple tasks.

\WN

Most definitely. As with everything, it is a matter of degree and
sequencing. Which types of tasks include a lower penalty?  When I ran
a team, we tried to keep an inventory of "filler tasks", that were
short and independent.  These often involved small updates to
documentation, reviewing, but were heavy on team management tasks.
For those that incur the highest penalty for blocks, (we typically
thought of design or debugging test cases) we made it a focus to
coordinate on the blockers. 

That is, having tasks ready to go is a good idea. However, I have
data that putting people on multiple projects is catastrophic. About
20\% of a persons time is consumed just *being on a project*.  A
weekly staff meeting consumes about an hour direct time, 15-30
minutes prep, and another 15-30 minutes transition.  That's 5\% of a
work week. I also have evidence that direct time on project tasks is
about 15\% of the work week, but at the team level, the coefficient
of variation for direct task time is about 25\%.  This was for teams
using TSP, who are presumably somewhat better than average with time
management.  (15 hours is about a 2.5 "Load factor" that some
agilests like to use to convert effort days to "ideal time")

\DJ

The fitted equation for WBS Estimate/Actual, based on summing
the values for the corresponding tasks, is:

\begin{equation}\mathit{WBS\_Actual\_mins}= 1.4 \mathit{WBS\_Estimated\_mins}^{0.94}\label{WBS_regression}\end{equation}

A slightly better fitting model can be obtained by including the number of
people working on the WBS, or the number of tasks in the WBS:

$\mathit{WBS\_Actual\_mins}= 1.5 \mathit{WBS\_Estimated\_mins}^{0.89}e^{0.08People}$, or

$\mathit{WBS\_Actual\_mins}= 1.7 \mathit{WBS\_Estimated\_mins}^{0.85}\mathit{Tasks}^{0.21}$

%

\subsection{Tasks}

\DJ

Tasks are the basic unit of work in TSP, with a WBS containing a
sequence of more or more tasks.  Figure~\ref{proj-WBS:fig}, right
plot, shows that 14\% of WBS contain a single task (e.g.,
\textsf{Misc}), and 53\% of WBS contain five or fewer tasks.  Within
the WBS of project 615, the most common task sequence is:
\textsf{Ident}, \textsf{Ident Inspect}, \textsf{Work},
\textsf{Inspect - Author}, \textsf{IT} (forming 22\% of WBS elements).

The following list shows the total number of each kind of task, by
phase name, for project 615.  The large role played by inspections in
TSP development is reflected in the frequency of this activity.

\begin{verbatim}
             Task Occurrences          Task Occurrences
 Inspect - Others        2031 Documentation         191
    Ident Inspect        1964        Design         169
             Misc        1171    Test Devel         167
            Ident        1153          Code         163
             Work        1037      Planning         160
 Inspect - Author        1024 Design Review         154
    Reqts Inspect         934          Test         153
               IT         848   Code Review         152
   Design Inspect         550    Postmortem         145
     Code Inspect         547       Compile         123
      HLD Inspect         371  Reqts Review          44
              HLD         204           \\N          16
         Int Test         195      Strategy           3\end{verbatim}

Figure~\ref{wbs-people:fig} shows information on the number of people
working on a WBS, per project, and the number of WBS per project.

\subsection{Work structure}

\DJ

Previous project task data I have analysed had a flat, linear,
structure; any higher level structure that existed in a project was
not present in the data.  I spent some time analyzing the CESAW data
assuming it had a flat structure, this was a mistake that led me down
various blind alleys.

The leaves of a WBS are self-contained elements of work involving a
sequence of tasks.  Successively higher levels in the WBS work
hierarchy specify higher level requirements, which are built from one
or more lower-level WBS elements; see figure~\ref{WBS-tree:fig}

\WN

A pattern for much of the work will be Create, Review/Inspect, Test.
These will apply to Requirements, High level designs, Code, and other
products like documents.  The Phase list should include the
overarching set of phases from which workflows will be derived. 

A development tasks normally includes design, reviews and
inspections, code, code review and code inspection, unit test. 

Requirements usually include some requirement's development, Builds
*usually* are a separate WBS and may include Build and integration,
and system test. 

Most code activities will include maybe (design, design review,
design inspection), code, code review, code inspection, and unit test.

Review is a personal review, inspection is peer review and usually
becomes 2-4 tasks. 

Teams have a lot of flexibility in setting up their workflows.  The
standard process phases help a lot to understand what is actually
going on.  They also have flexibility in their work habits.  This one
planned item was worked in two sessions, one of 81 minutes, the
(other the next day) of 131 minutes.  That's a little over an hour
and a little over two hours.  Those are very reasonable work times.
Neither work session recorded interrupt time, but for 1-2 hour
sessions, that is also reasonable.

\DJ

This explains one of the questions on my list, i.e., why does
\textit{CESAW\_Time\_fact} contain so many more entries than the Task
facts.

The data is now even more interesting. It is possible to look at how
individual tasks are split up over given intervals.

\WN

Yes, the time log is very interesting because you can see exactly
when direct work was performed. I was surprised I could not get the
community interested in using it because the data tells us a lot
about work habits.  We always coached that the sweet spot for work
sessions was a 90-120 min then a break. Actual direct time is seldom
more than 50\% of the work week, and anything above 35\% (14 hours)
is good, >16 hrs/week was outstanding, 20 hrs/week was best in class.
More than 50\% always was a flag for bad data.  

Essentially, two 2-hr work sessions a day (without interruption)
resulted in first-rate effort time.

Other flags include long sessions (our rule of thumb was 180 min)
without interrupt time. 

This needs additional context and explanation.  The phases names are
not intended to capture the *activity* as much as to highlight
rework.  The phase types are Creation, Appraisal, Failure, Other.
Creation phases (\textsf{Design}, \textsf{Code}) include only the
first draft by the author. This is expected to be the creation work,
everything else is correcting defects.  Test is a special case that
deserves additional discussion.

In test, there is the test execution time, and the defect fix/rework
time. Double entry accounting includes time in test along with Defect
fix times.

Defect fix time includes finding a defect (debugging, reviewing,....)
fixing the defect (some code/compile) and rerunning/evaluating test
results.

The difference between total test time and defect fix time
approximates the time a zero defect product would require.  
 
This is a systems engineering team performing many of the early
lifecycle activities.  Over the course of several projects, they
changed the high level process several times, but the early phases
were requirements development activities.

The sequence you show doesn't look like quite the right order, but
I'm sure I have that recorded in the fact sheets. Memory fails, so I
will have to look up the location.  The key idea is that they had a
Requirements Identification followed by a Requirements
development/documentation phase.  

Inspection is for code is somewhat separable because we recommend
(strongly) never inspecting for more than 2 hours at a time, and
proceed at a rate no more than 200 LOC/hr.  Larger packages, thus
would require multiple sessions.

\textsf{Inspect Author} is the rework the author must perform in
addressing the inspection findings.  This is well named for clarity,
unlike how inspection recorded for coding. 

For accounting purpose phases are a logical "waterfall". The
individual coding task is performed, reviewed by the author, then
send for inspection. All rework/recoding is recorded as an inspection
task by the author.  Code splits will record only how many sessions
the author required to complete the initial code draft. 

This may seem plodding, but they escaped something on the order of <1
def/KLOC new and changed code. Their best work was measured in two
digit defects/MLOC

\DJ

\begin{figure}
\begin{center}
\includegraphics{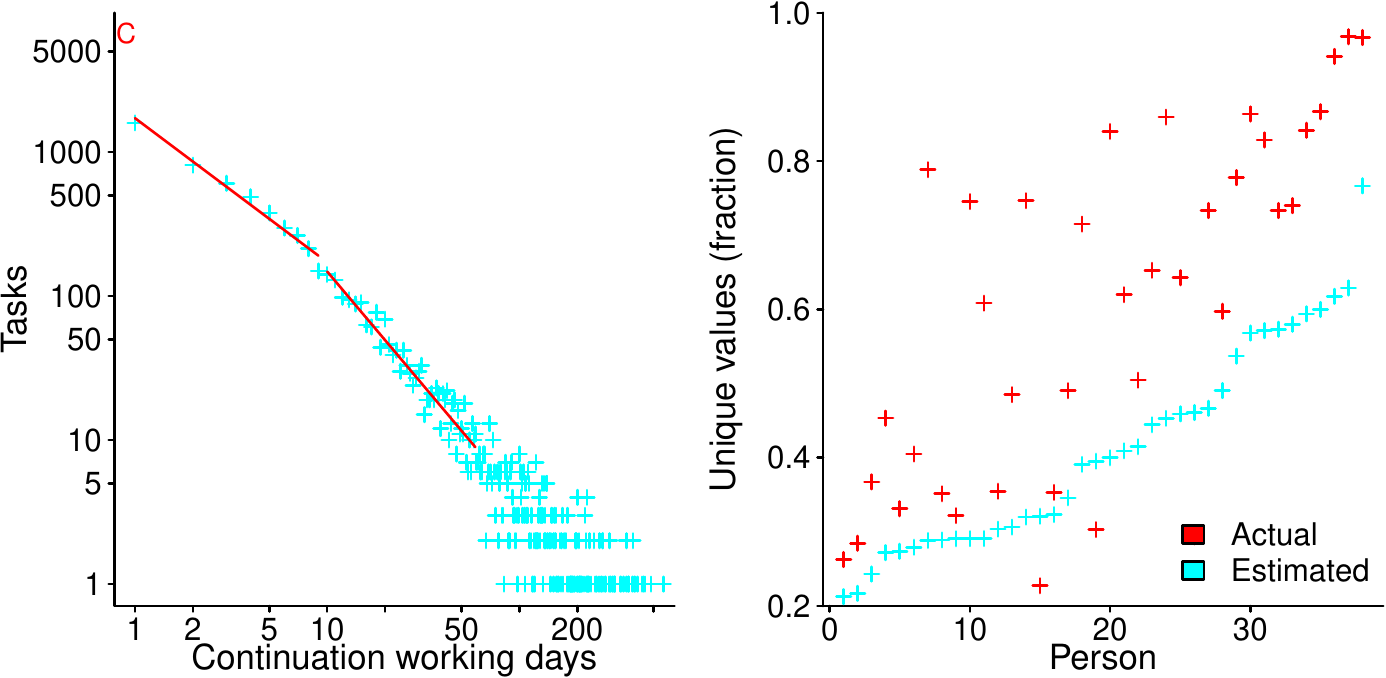}
\end{center}
\caption{Both plots are for project 615.  Left: Number of tasks that continue to be worked for a given number of working days after the first day, red C denotes all tasks completed on the day they were started; lines are fitted power laws with exponents of -1 and -1.6; right: Fraction of task estimates and actuals that are unique to each person who implemented at least 20 tasks, with people sorted by estimate fraction.}
\label{cont_days:fig}
\end{figure}

Work on some tasks was split across multiple sessions.  For instance,
the person involved may have taken a break, or had to pause while a
dependency was resolved.  When work on a task is split, there is a
separate row in the time log for each work session on the task.

When tasks are split across two sessions the most commonly involved
phases involve inspections; for several projects the most common
phases are: \textsf{Design Inspect} and \textsf{Code Inspect}.

How many tasks are not completed on the day on which they are
started?  Figure~\ref{cont_days:fig}, left plot, shows the number of
tasks whose end date was a given number of working days after the
start date, for project 615 (similar results are seen for projects
614, 617 and 95); red lines are fitted power laws with exponents of
-1 and -1.6.  The C, in red, shows the number of tasks completed on
the day they were started.

The change of slope in figure~\ref{cont_days:fig} happens at around
9-work days, i.e., 2 working weeks.  Is 2-weeks a special duration of
some kind?

It was not possible to fit a regression model connecting continuation
days to: planned time, number of work sessions, or day of the week.

\WN

Our guidance was that developers should try to complete 2-4 tasks per
week. By keeping tasks small, they can track work more accurately and
lower work in progress (WIP). We held weekly status meetings, so we
could expect some uncompleted work to show up as WIP. If WIP showed
more than a half week, that indicated a problem. We would look at the
open tasks to see if there was something we needed to do. If the task
went beyond a couple of weeks, this was almost usually a problem.
Sometimes it was just too big, or was stuck waiting.  We sometimes
tracked miscellaneous work in what we called a "bucket".  This was
something we wanted to track, but didn't correspond to a direct
deliverable, it was more a level of effort (LOE). Support tasks like
maintaining the revision control system, for example, would be
allocated some time in a metaphorical "bucket", from which time would
be drawn as needed. We often did this to maintain an historical
record of effort, so we could plan for it in the future.  This sort
of bucket was more a LOE than an earned value task.  But they needed
to be time boxed and closed, or they would mess up tracking by
increasing the noise to signal.

\DJ

\begin{figure}
\begin{center}
\includegraphics{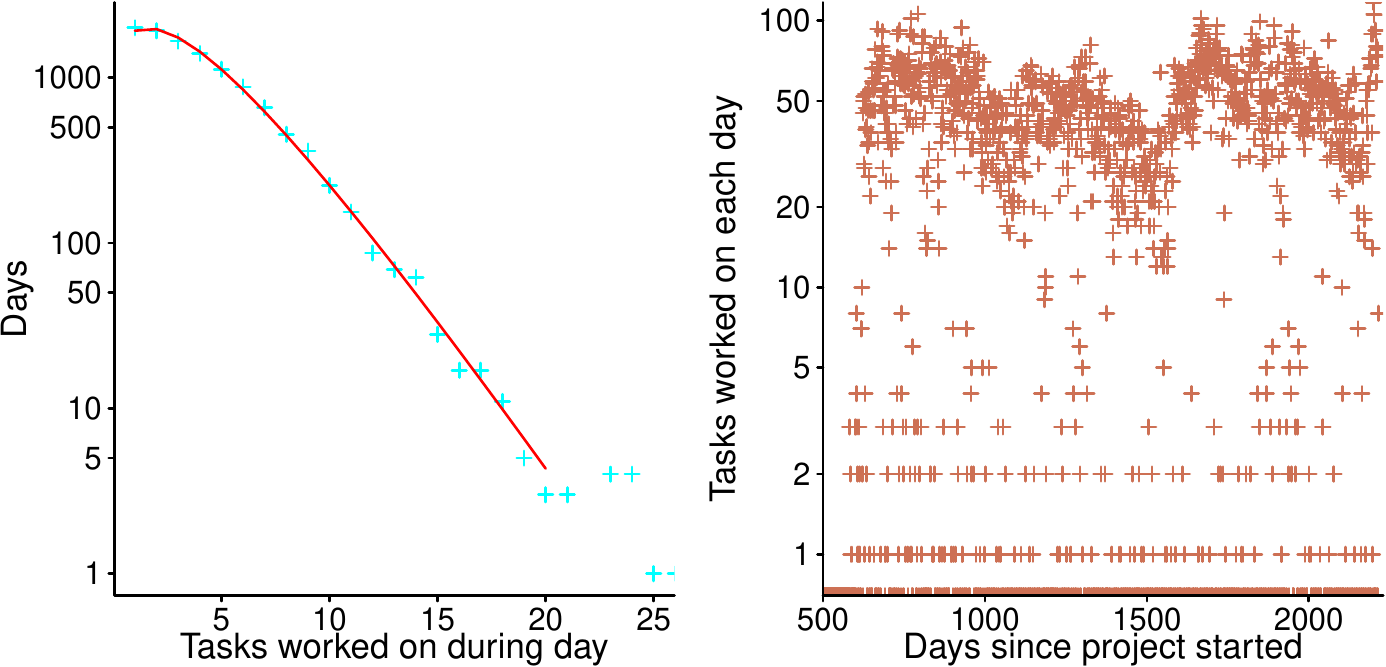}
\end{center}
\caption{Left: Number of days on which a given number of tasks were worked on during a day's work, for project 615, red line is a fitted zero-truncated Negative binomial distribution, right: Total number of tasks worked on each day of project 615.}
\label{zt-nb:fig}
\end{figure}

A person may work on more than one task per day.
Figure~\ref{zt-nb:fig}, left plot, shows the number of days on which
each person spent the day working on a given number of tasks, for
project 615; red line is a fitted zero-truncated Negative binomial
distribution.\footnote{Zero-truncated because zero values do not
occur.}

Including the number of sessions worked for each task,
$\mathit{num\_sessions}$, in a regression model updates
equation~\ref{first_regression} to give the following fitted equation:

\begin{equation}\mathit{Actual\_mins}= 3.1 \mathit{Estimated\_mins}^{0.55}\times\frac{\mathit{num\_sessions}^{0.75}}{e^{0.24\times(\mathit{day\_diff} = 0)}}\label{act-est-mod}\end{equation}

where: $\mathit{day\_diff} = 0$ equals $1$ when the task started and
completed on the same day, and $0$ otherwise.

This model explains 71\% of the variance in the data.  Similar fitted
equations were obtained for projects 95, 614, 617.

If completing a task requires multiple sessions, the weekends at
either end of a working week are a potential delimiting barrier.  Do
people aim to have in-flight tasks completed by the end of the week?

\begin{figure}
\begin{center}
\includegraphics{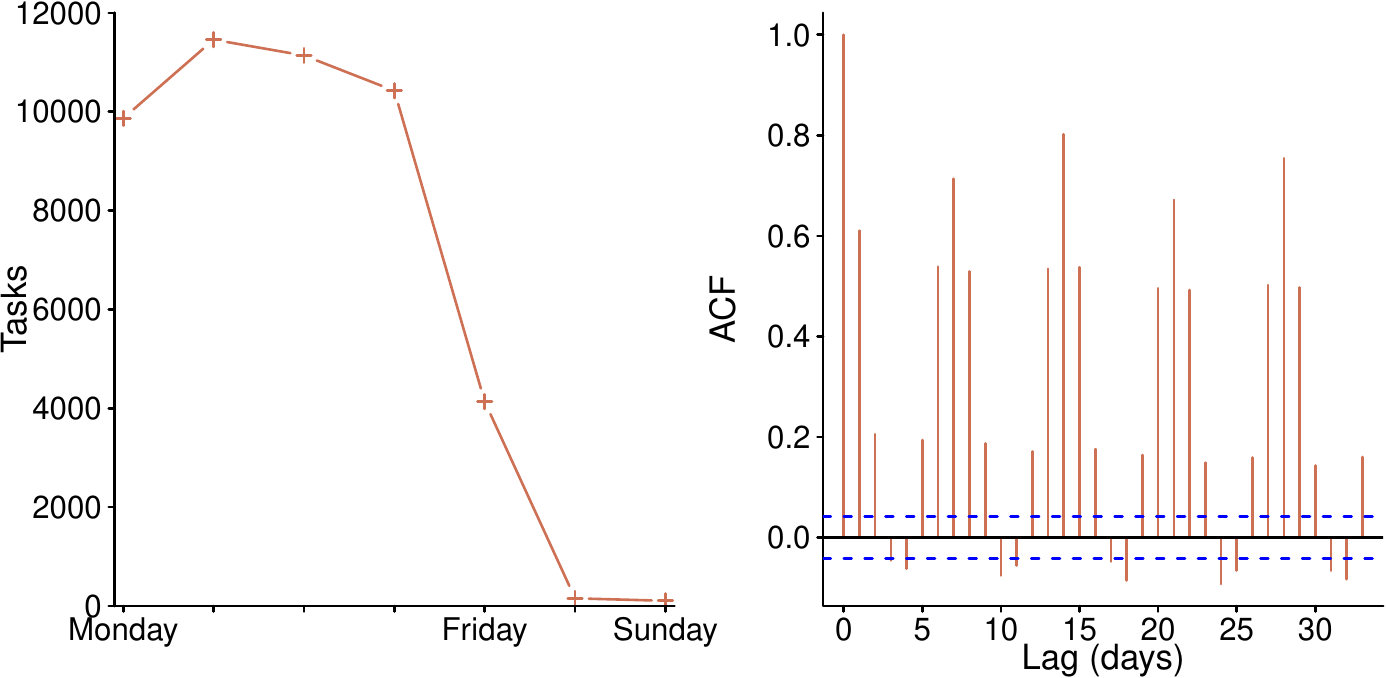}
\end{center}
\caption{Left: Total number of tasks worked on during a given day of the week, for project 615; right: Autocorrelation of the tasks worked on per day, for project 615.} 
\label{weekly-tasks:fig}
\end{figure}

Figure~\ref{weekly-tasks:fig}, left plot, shows the total number of
tasked worked on during a given day of the week, for project 615.

Figure~\ref{weekly-tasks:fig}, right plot, shows the autocorrelation
for the total number of tasks worked on per day, for project 615.
The recurring peak at 7-day intervals show that the number of worked
tasks on any day is highly correlated with the number of worked tasks
on the same day of the previous and following weeks.  The high
auto-correlation values either side of each peak implies a strong
correlation between the number of worked tasks on a given day, and
the previous day of the previous/next week and the following day of
the previous/next week.

\WN

The natural planning time frames were the day, the week, and the
"cycle". Developers thought in terms of a work day and would make a
plan of the day (PLOD) with the intention of trying to get something
done. The team planned weekly, so the teams and individuals built a
plan of the week (PLoW) during which they would reorder, add or
remove, and if necessary re-estimate their remaining tasks.  

The overall plan was broken into cycles, similar to sprints, where
the high level goals were detailed out for planning.  We found,
empirically, that a natural timescale for a detailed plan horizon was
12 weeks, (three months or a quarter). That is, for about 12 weeks,
we could manage the detailed plan to the quarterly targets.  After
that, plans tended to get so out of date that they needed to be
refreshed with a thorough replan. Some teams did this monthly, but we
tended to find quarterly was a good frequency to have the team step
back to revisit goals, priorities, risks, and long term strategy.

\vfill

\subsection{WBS staffing}

\DJ

What is the process used to decide who will be involved in the
implementation of a given WBS?

\WN 

There is the macro dimension between teams and the micro level within
teams.  Real projects are messy. I wrote short paper that listed some
common staffing patterns we observed \cite{Shirai_14b}, which found
various patterns in people working together. None of them are
surprising, but it's useful to understand some different work flow
structures involving teams and multiple teams. Sometimes they work in
parallel, sometimes in sequence by speciality, and in other cases
people came on during periods of intense work. 

\DJ

\begin{figure}
\begin{center}
\includegraphics{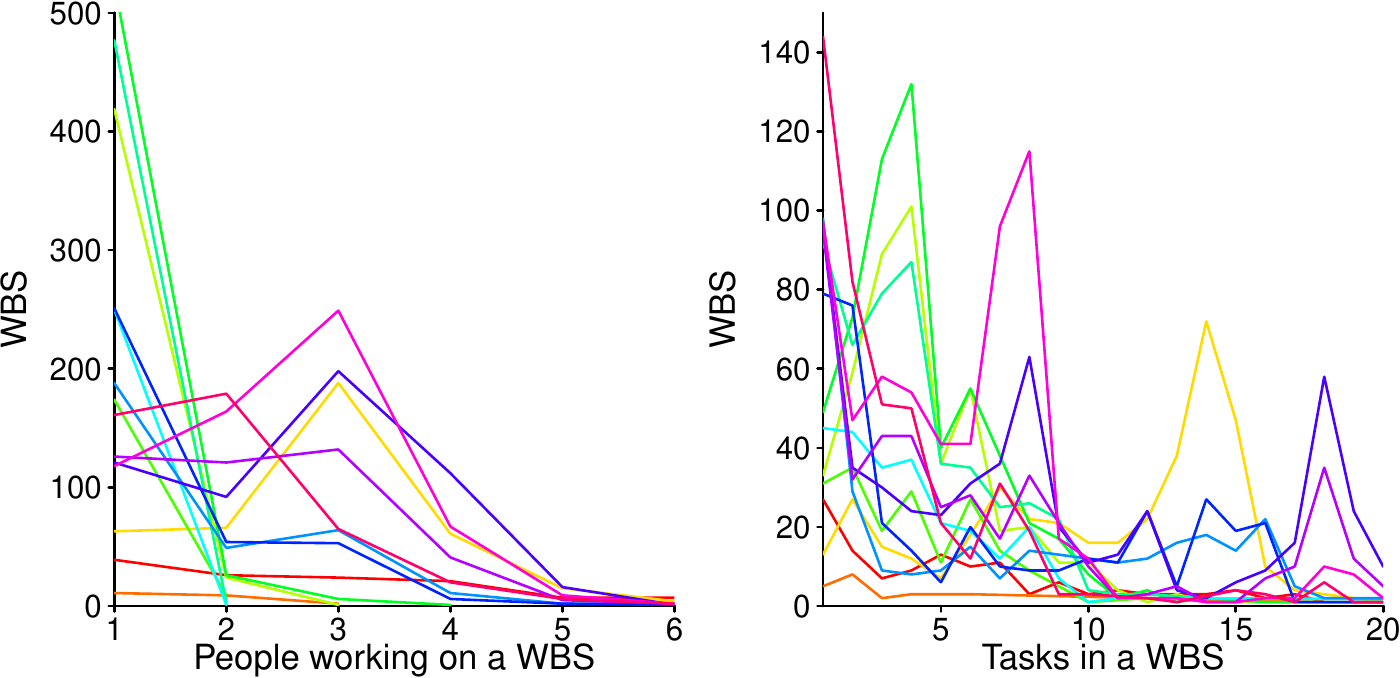}
\end{center}
\caption{Left: Number of WBS elements involving a given number of people, broken down by project (colored lines), right: Number of tasks per WBS by project (colored lines).}
\label{wbs-people:fig}
\end{figure}

Figure~\ref{wbs-people:fig}, left plot, shows the number of WBS
elements that involve a given number of people, broken down by
project (colored lines).  On some projects most WBS involve one
person, while on other projects a three-person WBS is the common case.

Figure~\ref{wbs-people:fig}, right plot, shows the number of WBS
elements that contain a given number of tasks, broken down by project
(colored lines).

\WN

That's an interesting view, and I'd never seen it presented that way
before. With Yaz Shirai, we looked at multi-project teams, and only
looked at people and time. That plot shows patterns on a single
project rather than multiple team projects. It's a pity this hasn't
been studied more, there may be some insights about why they work a
certain way and what is effective in different situations. 

\DJ

\begin{figure}
\begin{center}
\includegraphics{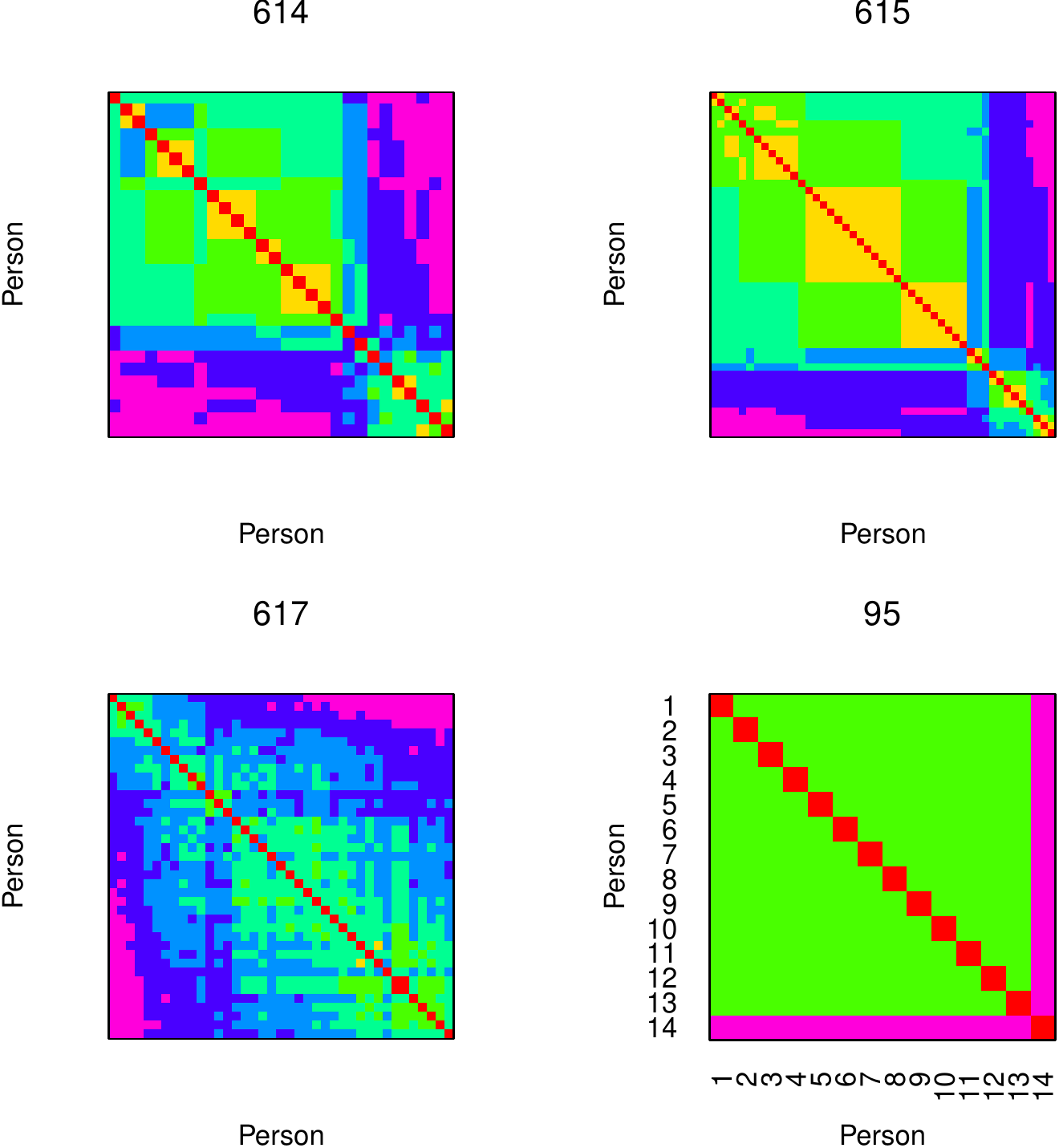}
\end{center}
\caption{Clustering of pairs of people based on number of times they worked together on a WBS, for projects 614, 615, 617 and 95.}
\label{wbs-seriation:fig}
\end{figure}

Within a project, how often do the people work together on a WBS?

Figure~\ref{wbs-seriation:fig} shows, for projects 614, 615, 617 and
95, a clustering of pairs of people working together on the same WBS,
based on the number of times they worked together on a WBS.

\section{Staffing}

\DJ

Companies are often concurrently implementing multiple projects, and
the implementation of a large system may be split into multiple
projects.

It is unlikely that the demand for staff on a new project will occur
just as a project is being completed.  One way of reducing the time
staff spend not earning income between projects is to have them work
on multiple projects, e.g., starting work on a new project while an
existing project is not yet nearing completion.

When projects are components of a larger system, it may be necessary
for them to be implemented concurrently; for instance, to allow a
common interface between them to evolve by agreement through
implementation experience.

\begin{figure}
\begin{center}
\includegraphics{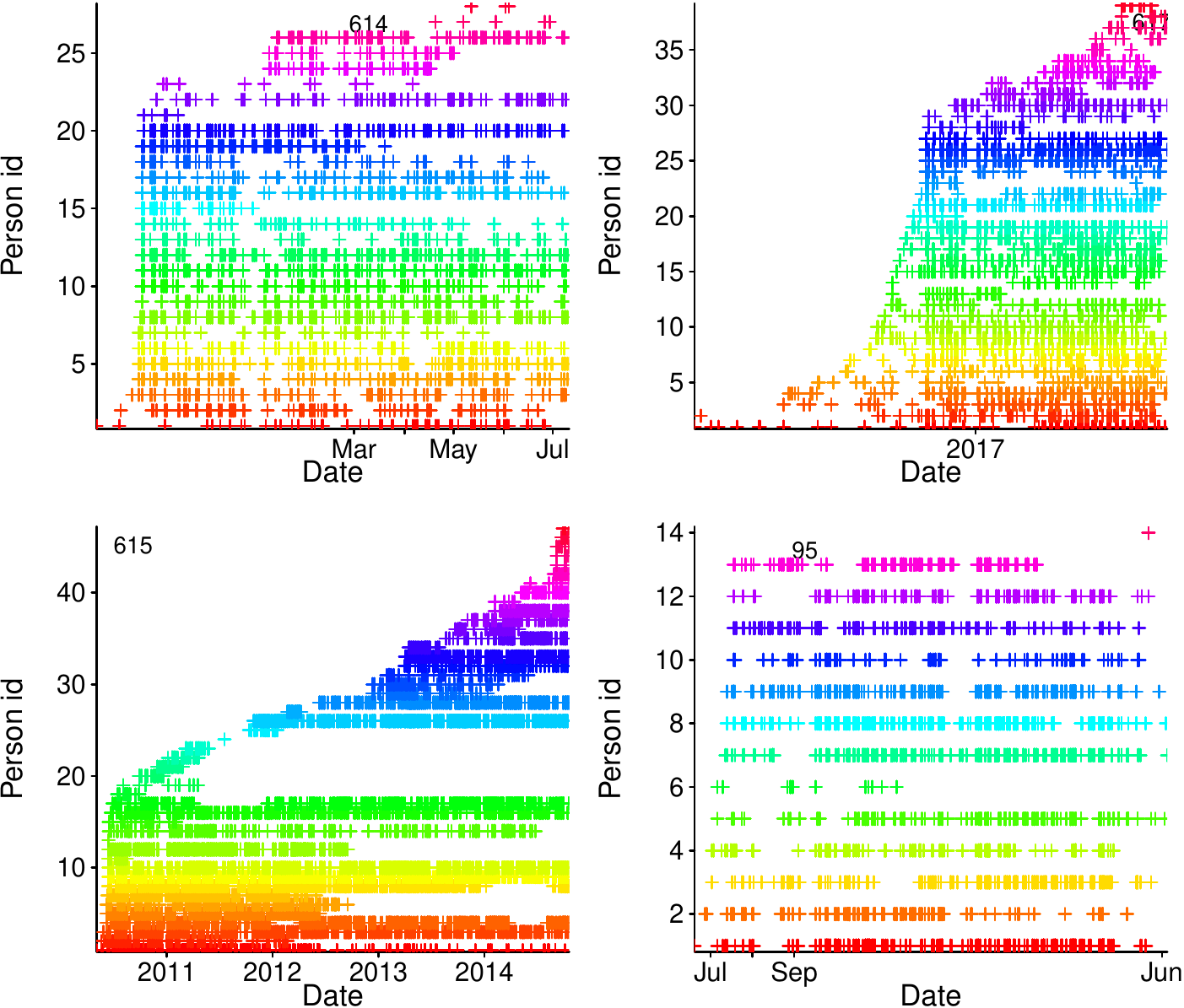}
\end{center}
\caption{Date on which each project member started working on a task, for projects 614, 615, 617 and 95, people sorted by date of first task.}
\label{person_timeline:fig}
\end{figure}

Figure~\ref{person_timeline:fig} shows the date on which each person
working on a project started each of their tasks, for projects 614,
615, 617, and 95; people are ordered on the y-axis by date of first
task.

The values appearing in the \texttt{person\_key} column, denoting
distinct people working on a project, vary across projects, i.e., it
is not possible to match up the same person working on different
projects.

\WN

Yaz Shirai and I looked at some common patterns and found 6 or that
were pretty common.

Some projects have cross-functional staff. The same people do all the
work.  Or sometimes, only the development work is closely tracked.
Both of these look like the same people from beginning to end.  In
other cases, specialists might come onto the team for a period of
time, for example requirements or test staff.  Then there were more
specialized teams. The requirements, architecture, or test may be
teams of specialists, The point is that teams had different ways of
organizing and staffing their work and there was no one size fits
all.  Another pattern we saw was a team or team members dropping in
during a period of high work load. 

\begin{figure}
\begin{center}
\includegraphics{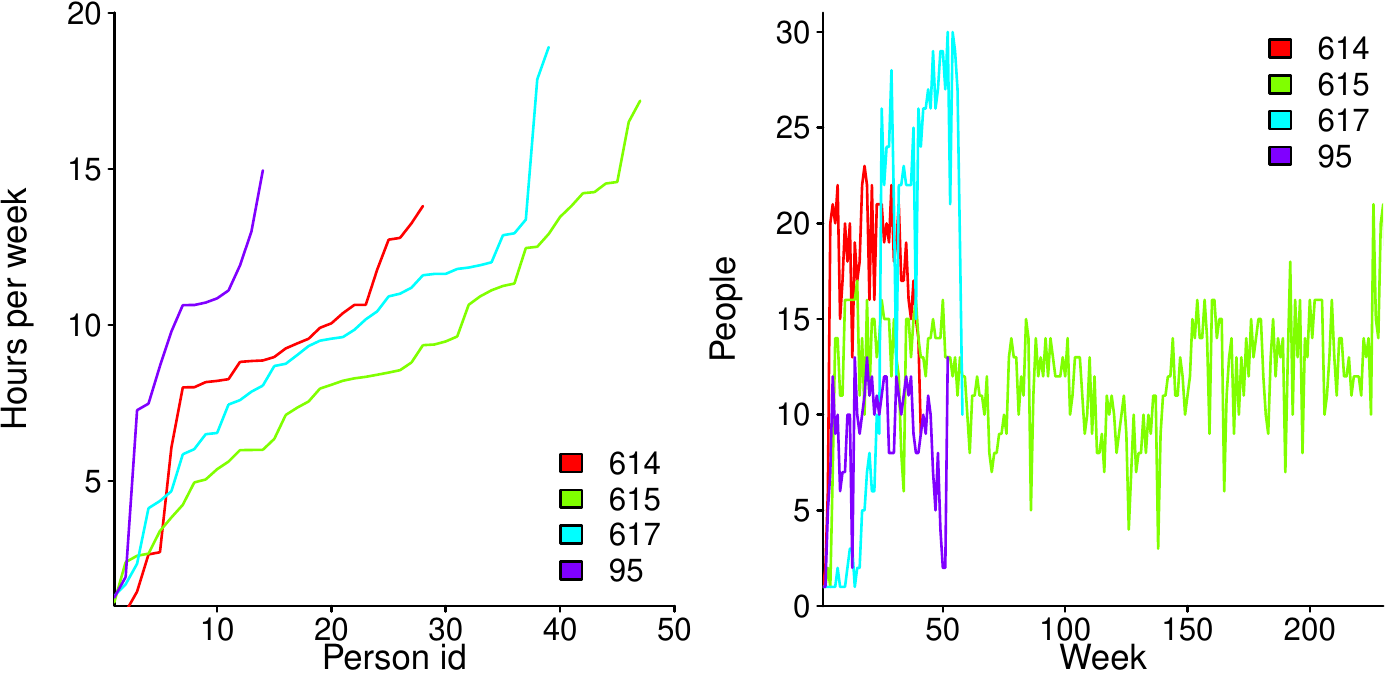}
\end{center}
\caption{Left: Average hours per week each person spent working on projects 614, 615, 617 and 95, people sorted by hours per week worked; right: Number of different people starting work on a task during the week.}
\label{person_percent:fig}
\end{figure}

\DJ

What percentage of their time do people spend working on a project?

Figure~\ref{person_percent:fig}, left plot, shows the average number
of hours worked per week (i.e., total hours worked divided by number
of weeks between starting work on the first and last tasked worked)
by each person who worked on projects 614, 615, 617, and 95.  People
are sorted in hours per week order, and person numbers do not
correlate across projects.

How many people are actively working on a project, over time?

Figure~\ref{person_percent:fig}, right plot, shows the number of
different people starting work on a task during each week.

Reading Humphrey's book \cite{TSP-2000} I did not see anything about
the distribution of people across projects.  Is there any informal
practice?

\WN

There are a lot of opinions on how it should be done, but very little
empirical work quantifying what people actually do.  My observation
is that smaller companies are more likely to be cross-functional and
bigger places have more specialists.  Bigger companies also had more
teams, so it was easier to just move work between teams rather than
add people to an existing team.  We strongly encouraged keeping teams
stable and had strong evidence that someone should be on only one
team.  The overhead for being on a team was about 20\% of total time.
It was tempting to "phantom staff" by spreading people. We showed
from time logs that a second projects didn't create two half-time
people, but two third-time people. A third project usually meant you
had someone who only attended meetings for two teams or who ignored
two teams. 

\section{Estimates and actuals}

\DJ

A wide range of cognitive and social factors have been found to
influence human time estimates\cite{Halkjelsvik_18}.  When giving a
numeric answer to a question, people sometimes choose to give a value
that is close to, but less exact, than what they believe to be true;
values may be rounded towards preferred values, known as
round-numbers.  Round-numbers are often powers of ten, divisible by
two or five, and other pragmatic factors\cite{Jansen_01}; they can
act as goals\cite{Pope_11} and as clustering
points\cite{Sonnemans_06}.

\begin{figure}
\begin{center}
\includegraphics{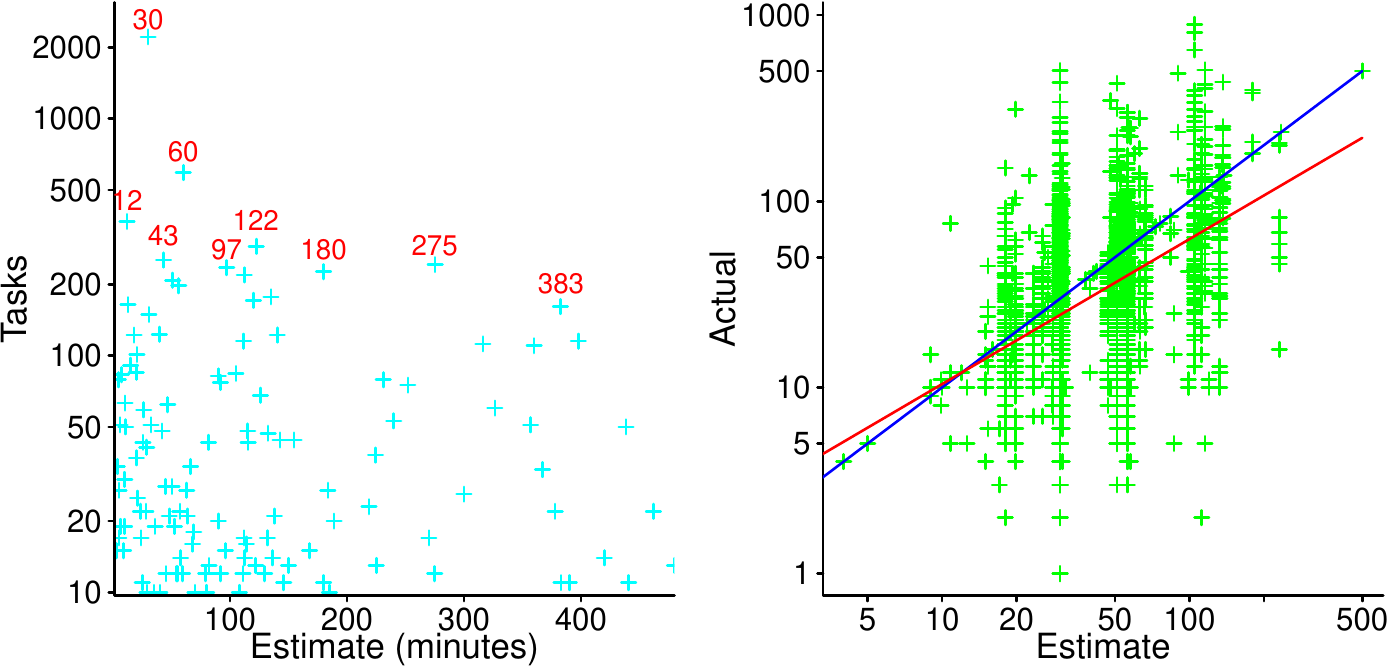}
\end{center}
\caption{left: Number of tasks estimated to take a given number of minutes, for project 615, with some peak values highlighted (red numbers); right: Estimates and actuals for \textsf{Inspect - Others} tasks in project 615, with fitted regression line (red) and $\mathit{actual}=\mathit{estimate}$ line (blue).}
\label{num_ests:fig}
\end{figure}

Analysis of task estimation data from other projects has found that
frequent use of round-number values is common, e.g., 30 and 60
minutes.  The frequent appearance of round-numbers in actual task
times has not been as common.

Figure~\ref{cont_days:fig}, right plot, shows the fraction of unique
estimate and actual values, for each person who worked on at least 20
tasks, for project 615.

\subsection{Estimates}

\DJ

While this section focuses on the estimates made for project 615,
similar patterns can be seen in the estimates made for other projects.

Figure~\ref{num_ests:fig}, left plot, shows the number of times a
task is estimated to require a given number of minutes, for project
615.  While round-numbers such as 30 and 60 are very common, some
surprising (to me) numbers are also common, e.g., 12.  Many of these
surprising estimate values are for inspection tasks.

Figure~\ref{num_ests:fig}, right plot, shows Estimate against Actual
for the \textsf{Inspect - Others} tasks of project 615; the red line
is a fitted power law, and the blue line shows
$\mathit{Actual}=\mathit{Estimate}$.  The large number of estimates
clustering around certain values is visible in the concentration of
points forming vertical shafts.

Adding the (anonymous) identity of the person doing the inspection to
the regression model (i.e., equation~\ref{act-est-mod}) significantly
improves the fit.  The fitted model shows that estimate accuracy
varies by a factor of more than 10 across inspectors.

\begin{figure}
\begin{center}
\includegraphics{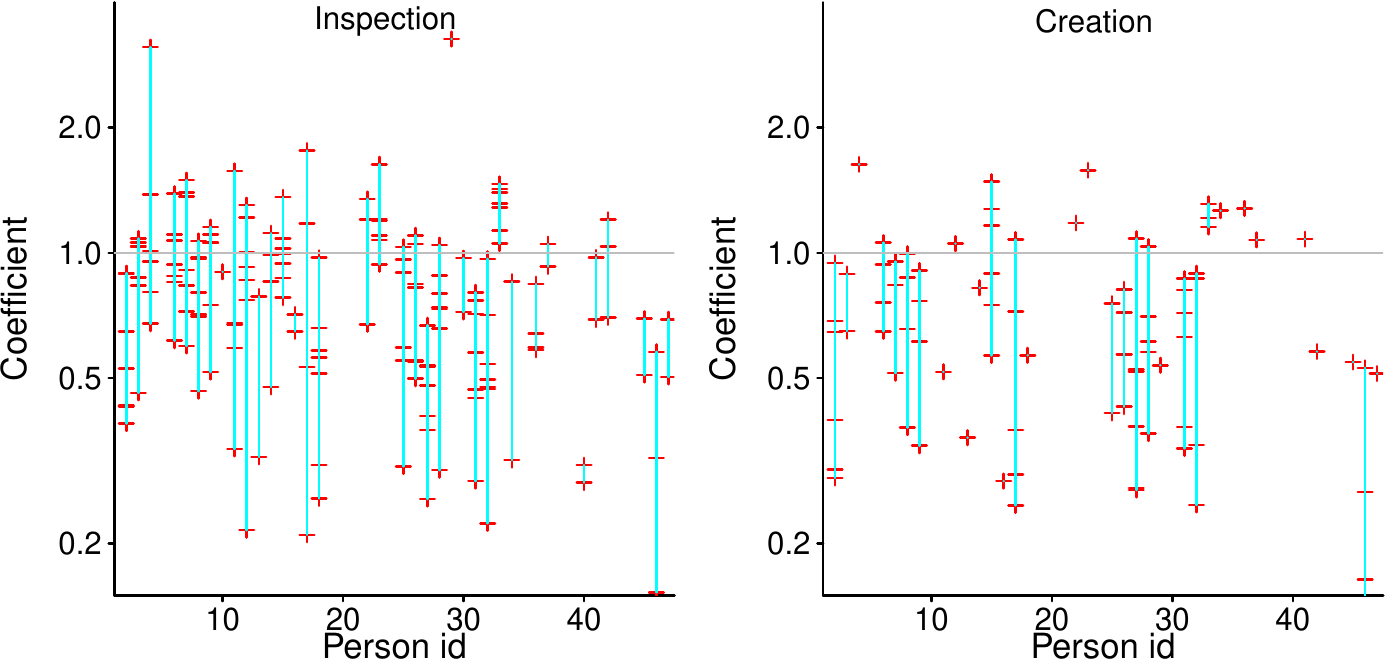}
\end{center}
\caption{Individual multiplicative factor, for each person who made at least five estimates, combined into equation~\ref{act-est-mod}; below grey lines shows an overestimation trend, and above grey line an underestimation trend, for the seven possible inspection tasks (left) and seven possible creation tasks (right), for project 615.}
\label{person-insp:fig}
\end{figure}

Figure~\ref{person-insp:fig} shows the multiplicative factor added
into equation~\ref{act-est-mod}, for each individual who made at
least five estimates on project 615.  Points below the grey line
indicate overestimation, compared to group average, while points
above the grey line indicate underestimation.  The left plot is based
on the seven possible inspection tasks (denoted by distinct red plus
symbols for each person), and the right plot is based on the seven
possible creation tasks.

Why is there so much variation in the relative accuracy of individual
estimates?

Possible reasons include: risk tolerance, implementing tasks that
involve more/less uncertainty, inexperience, and factors external to
the project that have a greater need of cognitive resources.  Data
analysis can highlight a behavior, uncovering likely causes is down
to project management.

\WN

Inspection tasks have two recommended constraints.

1) the inspection rate should be less than 200 LOC/hr or about 4
pages an hour for typical documents,

2) the size of the inspection needs to be limited so that inspectors
don't spend more than an hour or 90 minutes in a sitting.

Based on size, the effort is typically:
$\mathit{Effort\_Estimated}= \mathit{Size}\times\mathit{Historical\_Rate}$

Tasks should be estimated at the leaf WBS. A WBS element has Plan
items which are tasks. A process (workflow) is applied to a WBS
component, and the activities are each allocated a percent of the
total effort.  This way, while the development of WBS component (for
example a story) may take a couple of calendar weeks to complete, but
would have separate tasks for design, design review, design inspect,
code, code review, code inspect, (compile), and test. 2 to 4 of tasks
should be completed in a week. This provides frequent feedback on
progress and limits the work in progress to about half a week's worth
of effort. 

\textsf{Inspect - Others} refers to the peer inspectors on a
requirement. The owner is responsible for collating and disposing of
all comments and IIRC, assigns this to *Inspect*.

The WBS in the task list should contain leaf elements. I expect this
distribution to be somewhat constrained. As we go up the tree, we
will likely find some distribution of number child nodes. One use
might be to estimate the planning effort required to prepare a
backlog. I've also seen instances where design re-work was required
to decompose large components so that work could be partitioned and
allocated. Without partitioning, a single developer could not
realistically have achieved the desired schedule.

Based on size, there may be a minimum amount of planning required to
get to some number of implementable pieces. There is likely a
realistic limit into how many pieces a component can be subdivided.
This may be useful to know.  One could probably make an educated
guess about the number of WBS levels required. This in turn may
provide a more principled way to estimate staffing, schedule, and
cost for requirements development, architecture, and design work.
Benchmarks might be a reality check that the planning is adequate, or
overdone.

\subsection{Actual time}

\DJ

\begin{figure}
\begin{center}
\includegraphics{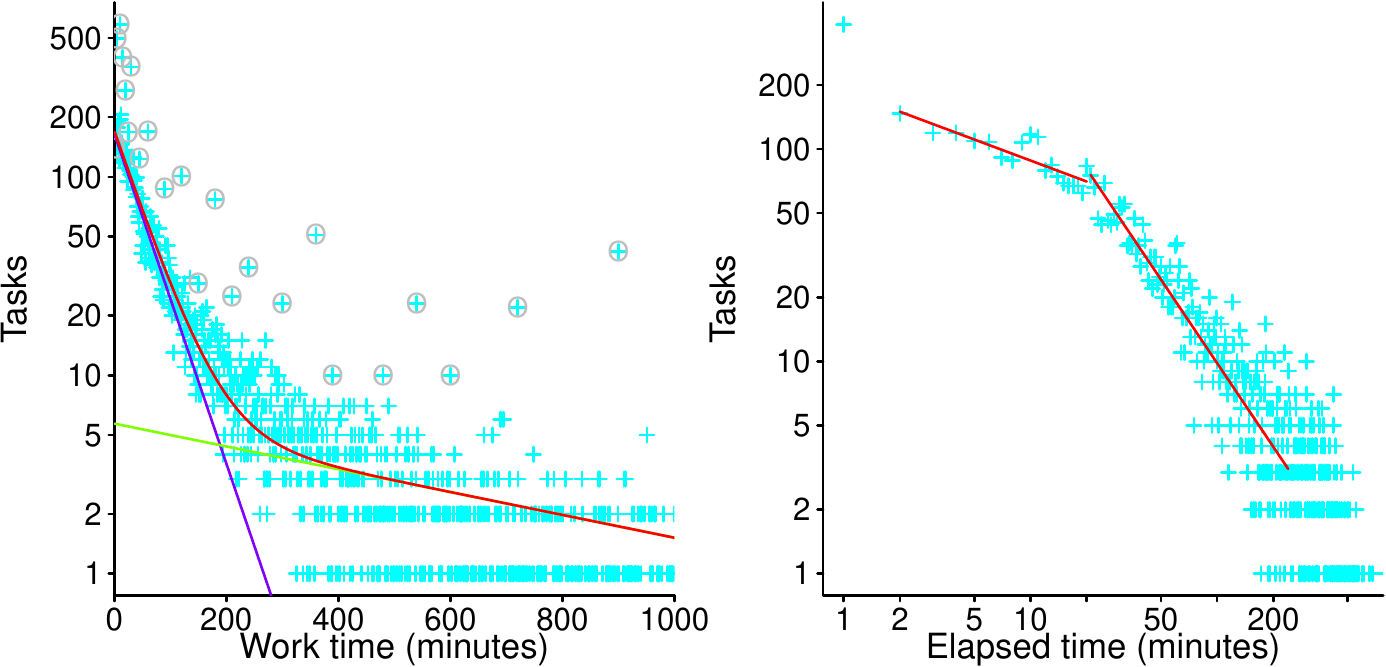}
\end{center}
\caption{Both plots based on data from project 615.  Left: Number of tasks worked on for a given amount of actual time, red line is a fitted bi-exponential (components in purple and green), grey circles specify spike values (which were excluded from the fitted model);  right: Number of tasks completed on the same day taking a given elapsed time, with two fitted power-laws (exponent up to 20 minutes is -0.3, and -1.3 after).}
\label{task_freq:fig}
\end{figure}

Methods of measuring the actual time taken to complete a task include:

\begin{itemize}
\item the amount of time spent working on the task.
Figure~\ref{task_freq:fig}, left plot, shows a fitted bi-exponential
(red line) to the data, for project 615, with spike values (circled
in grey) excluded from the fit; purple and green lines are the
components of the bi-exponential,

\item elapsed time between starting and ending a task, which may
include intervals where work on the task is stalled, waiting for some
event to occur.  Figure~\ref{task_freq:fig}, right plot, shows
elapsed time for tasks completed on the day they were started, for
project 615; red lines are fitted power laws (the break is at 20
minutes),

\item time spent working on the task, plus time consumed by work
interruptions; see figure~\ref{interrupt_freq:fig}.
\end{itemize}

In some cases, people may work hard to complete a task within the
estimated work time, or may work more slowly to use up the available
time.  Parkinson's law claims that work expands to fill the time
available.  Both behaviors predict a peak at the point where actual
time equals estimated time.

\begin{figure}
\begin{center}
\includegraphics{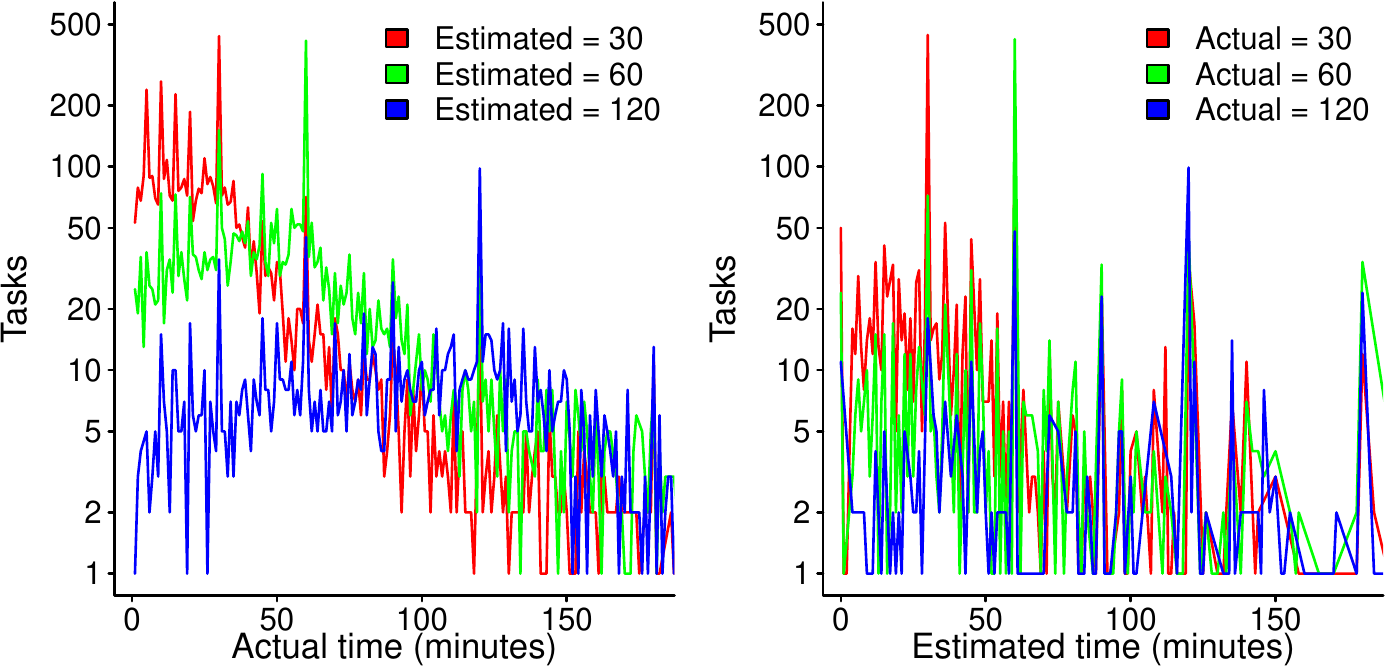}
\end{center}
\caption{Both plots are based on data for all projects.  Left: Number of tasks requiring a given actual time for the same amount of estimated time; right: Number of tasks estimated to take a given amount of time, having a given actual time to implement.}
\label{act_4_est:fig}
\end{figure}

Figure~\ref{act_4_est:fig} for all projects shows: left plot, the
number of tasks having a given actual time when a given amount of
time was estimated (e.g., 30, 60 or 120 minutes); the right plot
shows the number of tasks having a given estimated requiring a given
amount of actual time to implement.

Both plots show that the main peak, in number of tasks, occurs when
actual equals estimated time.  There are also smaller peaks at
round-number values less than and greater than the estimated time.

\subsection{Work interruptions}

\DJ

Work on a task may be not be continuous, e.g., input from other
activities may interrupt work, or a person may take a break.  The TSP
records interrupt time, and the splitting of a task across work
sessions.

\WN

A continuous work session to complete a task was the exception rather
than the norm. We found that working more than a couple of hours at a
time tended to be hard to sustain, requiring a short break. For
coding or testing a 5-10 minute break was often enough. But while
debugging was almost hypnotic, intense work such as design or
inspection was so physically demanding that more than a couple of
hours in a day was the max someone might be able to do, Since normal
tasks might take 5-10 hours, most required several work sessions,
often across multiple day. Moreover, there are a lot of other things,
like team meetings, support work, filling out the time sheet, and so
forth that get in the way. It was normal for only 10 to 20 hours per
week to go towards tracked tasks. For this reason, wall clock, or
calendar duration, had a weak correlation with direct effort.  One of
our productivity tricks was to have a team "quite time" or "focus
time", two or three hour blocks several times a week during which
there would be no interruptions. 

All this is pretty straight forward to track with the Process
Dashboard. During the plan of the day, you just place your top task
at the top of the stack, and maybe make sure the next two or three
are in sequence. The Plan of the Week should have these all ready.
The active task has an on/off checkbox with a timer. We used it like
a stopwatch. The task time only accumulates while the timer is on and
placed this in the log.

\DJ

How often is work on a task interrupted, and what is the duration of
interruptions?

The CESAW data contains the total number of interrupt minutes (i.e.,
the \texttt{time\_log\_interrupt\_minutes} column), but no count of
the number of interruptions.  The project average for percentage of
work sessions interrupted is 89\% (sd 6.9\%).

\begin{figure}
\begin{center}
\includegraphics{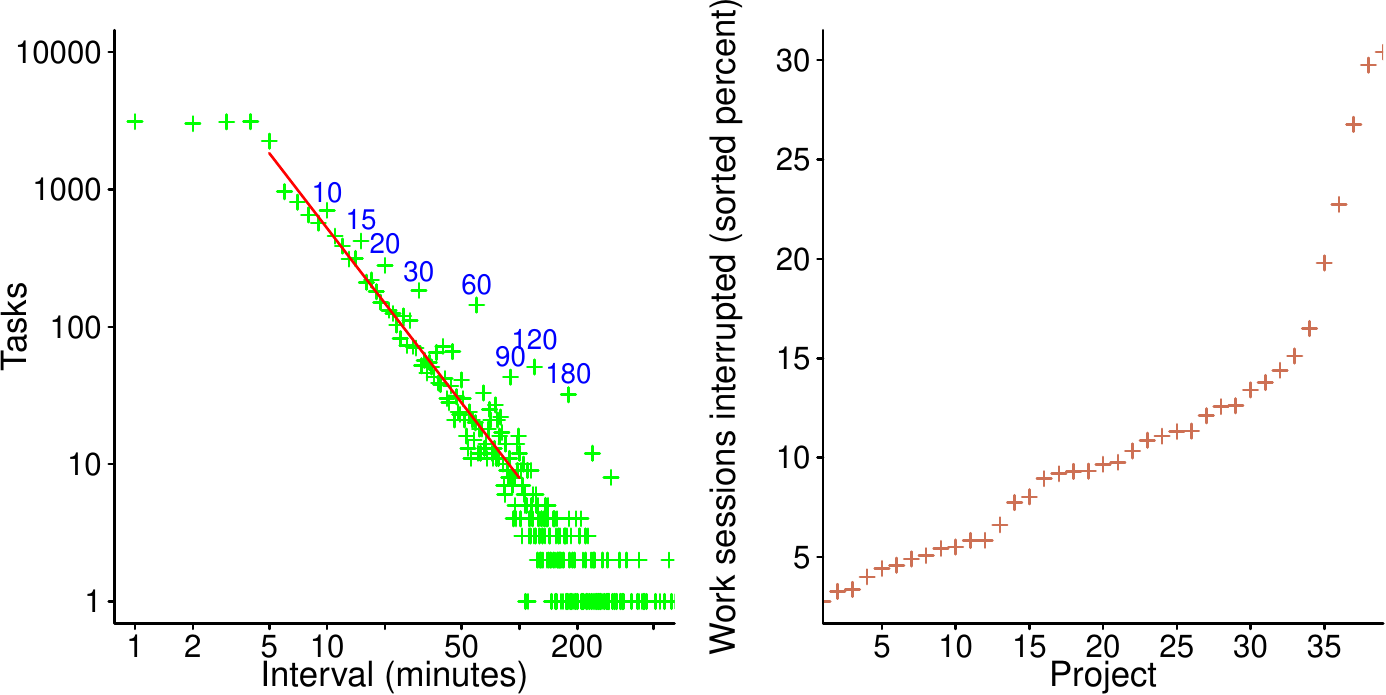}
\end{center}
\caption{Left: Number of task work sessions whose interruptions consumed a given amount of actual time; red line is a fitted power law of the form: $\mathit{freq} \propto \mathit{interval}^{-1.8}$, for all work sessions excluding spikes at round-numbers (some highlighted in blue); right: percentage of work sessions experiencing at least one recorded interruption, sorted by percentage.}
\label{interrupt_freq:fig}
\end{figure}

Figure~\ref{interrupt_freq:fig}, left plot, shows the number of task
work sessions having interruptions that consumed a given number of
minutes; the red line is a fitted power law having the form:
$\mathit{interrupt\_freq} \propto \mathit{interval}^{-1.8}$
(round-numbers that 'spiked' were excluded from the fit).  The right
plot shows the percentage of work sessions experiencing at least one
interruption, sorted by project.

The column \texttt{task\_actual\_time\_minutes} in
{\itshape CESAW\_time\_fact.csv} does not include interrupt time.
Including interrupt time in the regression models finds that its
impact, across projects, varies between 3-20\%.

Work on a task may be split across multiple work sessions (see
figure~\ref{cont_days:fig}), with 45\% of all tasks completed on the
day they are started.  A small percentage of tasks are worked on
during two sessions on the same day; figure~\ref{elapsed-time:fig},
right plot, shows the number session intervals having a given length
(in 15-minute bins).

\subsection{Impact of practice on accuracy}

\DJ

Through performing some activities, people get better with practice.
There is an ongoing debate about whether changes in performance,
through practice, are best fitted by a power law or an
exponential\cite{Heathcote_00}; in practice there is little
difference in the fitted curves over the range of interest.

The analysis of estimation accuracy in the SiP dataset
\cite{Jones_19a} found that developers did not improve with the
number of estimates made, and it was hypothesized that developers
prioritized learning to perform tasks more effectively (rather than
learning to improve estimate accuracy).

Adding a variable containing the $\log$ of the relative order of when
an estimate was made (for creation phases only), by each individual
developer (who made more than ten estimates), to regression models
similar to equation~\ref{act-est-mod} failed to improve the fit (for
projects 614, 615, 617 and 95).

\begin{figure}
\begin{center}
\includegraphics{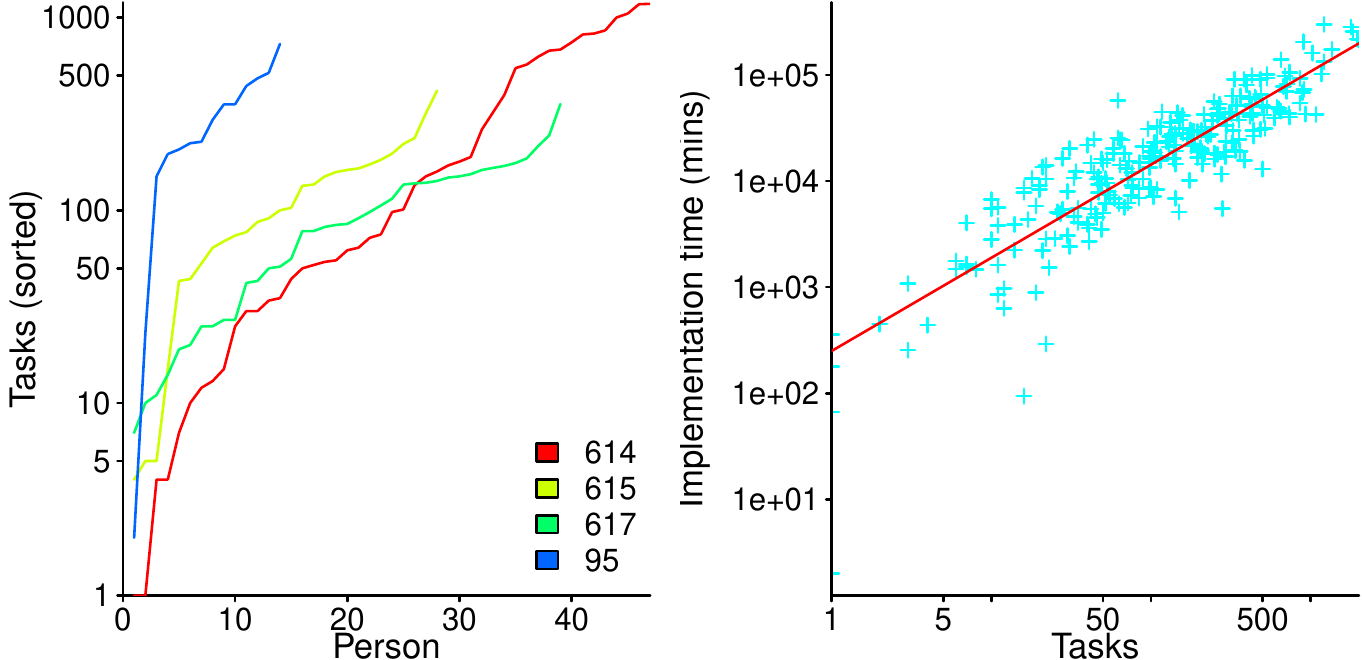}
\end{center}
\caption{Left: Number of tasks (in sorted order) implemented on projects 614, 615, 617 and 95, by each individual working on it; right: For all CESAW projects, the number of tasks each individual implemented against the total implementation time for those tasks, red line is a fitted power law with exponent 0.88.}
\label{task_person:fig}
\end{figure}

Figure~\ref{task_person:fig}, left plot, shows the number of tasks
performed by each person working on a project.  The right plot shows
for each of the 247 people working on the CESAW projects, the number
of tasks implemented, and the time taken to implement those tasks
(the red line is a fitted power law with exponent 0.88).

How were individual developer estimates evaluated when they were part
of a team?  Was everybody held equally responsible for accuracy?

Which factors and incentives influence the thinking process of the
person making an estimate?

It would not be cost-effective to spend more time estimating than it
is likely to take to do the job, and time spent estimating will be a
fraction of the likely estimation time.  One possible reason why
short duration tasks tend to be underestimated is that the person
making the estimate does not spend enough time studying the task to
notice the potential pitfalls; intrinsic optimism holds sway.

Staff incentives are driven by the work context, and may be
experienced differently by management and individuals.

\WN

Developers were expected to manage their own estimation. By and
large, they were making work commitments, so it was their
responsibility to make a reasonably accurate estimate.  Estimation
method varied by team, and we don't always have data for a basis of
estimates in the SEMPR. In some cases, they used a PSP workflow, and
this automatically used PROBE \cite{DSWE-Humphrey}.  High level
estimates usually involved a size estimate and an historical
development rate. Individuals were expected to apply their own rates
for work they were assigned.  Tasks were generally spread across
phases using the historical percentage time in phase.

\subsection{Wall clock time}

\DJ

What are the characteristics of daily work activities, relative to
time of day?

\begin{figure}
\begin{center}
\includegraphics{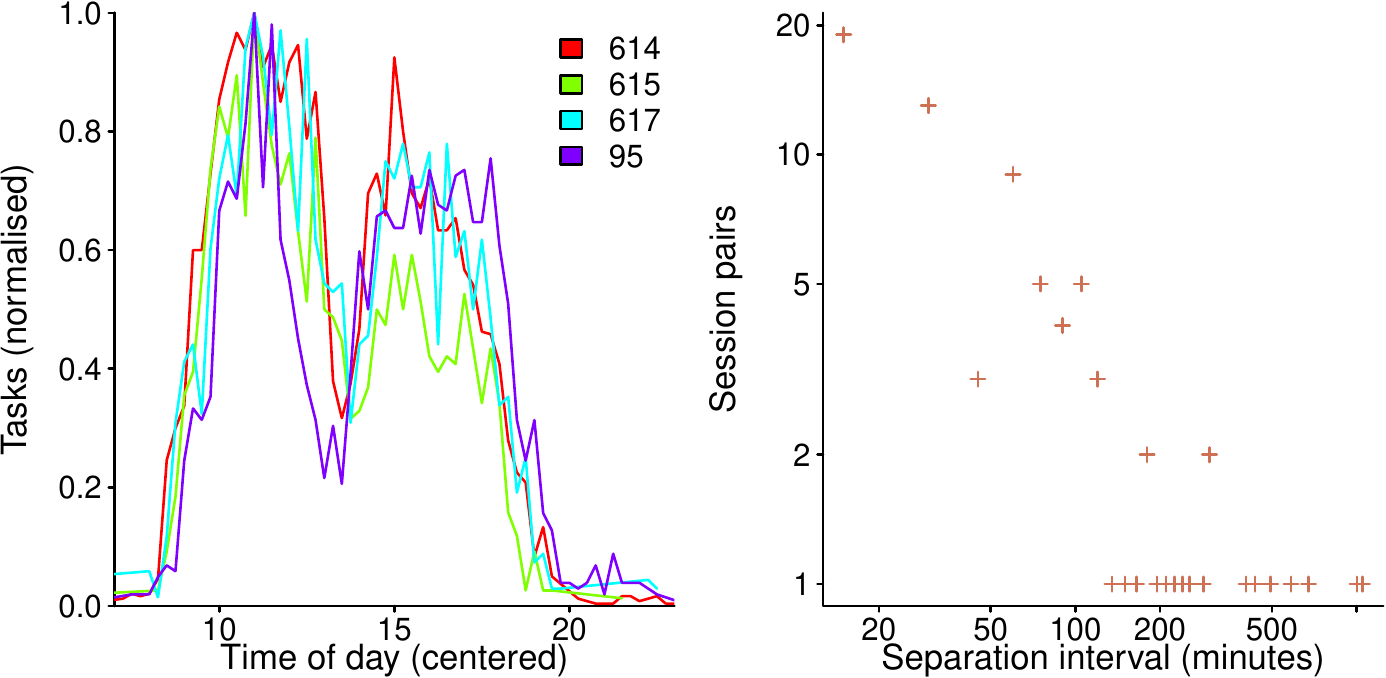}
\end{center}
\caption{Left: For tasks started and completed on the same day, number of tasks starting at a given centered time of day, within a 15-minute window for projects 614, 615, 617 and 95; right: For two sessions, of the same task, starting on the same day, the number of session intervals of a given length (in 15-minute bins).}
\label{elapsed-time:fig}
\end{figure}

Figure~\ref{elapsed-time:fig}, left plot shows the number of tasks on
which work started, within 15-minute intervals, for a normalised time
of day.  The time of day has been centered to allow comparison
between projects (the times given in the time log columns of
{\itshape CESAW\_time\_fact.csv} were converted from project local
time to US Mountain Standard Time).

\WN

For this team, typical working hours (before COVID-19) are something
like 0700 to 1700 Monday-Thursday (9-hr days w + 1hr lunchtime), and
an eight-hour day every other Friday (with the other Fridays off).
This is called a 'Flex' schedule, with every two weeks having 80 work
hours. That was similar to my work schedule when I worked at the lab. 

The 'core' hours are something like 0900-1400 so that people can come
in early or late but should be covering the 'core' times of the day
so that they are available for meetings. Personally, I used the early
start as my focus time since I know there would be no meetings
between 7:00 and 9:00.

The dashboard stores timestamps based on standard unix time. I
imported the data on a computer using US Eastern Time. This is where
things can get a little strange looking with the data. My computer
system always knows what time zone I am using.  When I imported a
team's data into the SEMPR, the time typically is stored in the
database using the time in my time zone.  Europeans will look like
they are starting in the very early morning, like 3:00 AM, while US
west coast teams will seem to start several hours later than an
Eastern Time Zone team.  

\DJ

Figure~\ref{elapsed-time:fig}, right plot, shows, for tasks worked on
during two sessions on the same day, the number session intervals
having a given length (in 15-minute bins).

\section{Round numbers}

\DJ

When giving a numeric answer to a question, people sometimes choose
to give a value that is close to, but less exact than, what they know
to be true.  Values may be rounded towards a preferred value, known
as a round-number.  Round-numbers are often powers of ten, divisible
by two or five, and other pragmatic factors\cite{Jansen_01}; they can
act as goals\cite{Pope_11} and as clustering
points\cite{Sonnemans_06}.

\WN

This is why we preferred T-shirt size estimation. We trained and
coached to avoid direct numeric estimates by using the T-Shirt size
method, Very Small, to Very Large.  

We converted the T-shirt size into cardinal numbers by calibrating
with historical data.  The T-shirt size tables had an average sizes
for each size bin.  We've found that size is log normally
distributed, so we took advantage of that observation.  We built the
size tables by log transforming historical size data so that it
looked roughly like a Normal (Bell) curve, then segmented it into
size bins based on standard deviations from the mean.  The actual
(untransformed) size ratio between bins was usually about 2.5.  

This is similar to how agile teams using story points a Fibonacci
sequence, but the biggest difference is calibrating a parametric
distribution to actual historical sizes.

\vfill

\subsection{The impact on accuracy}

\DJ

Picking round-numbers, rather than potentially more accurate values,
has the advantage that managers or clients may find round-numbers
more believable. Management/client influenced estimates are a
perennial problem.  Getting developers out of a round-number mindset
may not be useful if management/clients think in round-numbers.

\WN

That's an interesting perspective.  We had a Top-Down estimate before
work began, but the develops did their own bottom up estimate.
Aggregating the bottom up has at least the appearance of being more
precise. 

Of course, round numbers can be both inaccurate (biased) and
imprecise.  We tried to avoid bias through calibration with actual
data. A medium was a medium, and a large was a standard deviation
larger.  If the bins were centered and wide enough for people to
distinguish, we could have imprecise (but unbiased) point estimates,
but the central limit theorem drove big picture precision. This
relies on some parametric assumptions that turn out to work in most
cases. 

BTW, we could also look at a plan to see if the estimates were
balanced around medium.  This was often used to check for bias during
the planning. 

\DJ

Did people actually use the T-Shirt technique to make estimates?  The
task estimates have too many distinct values to be T-Shirt based, see
Figure~\ref{cont_days:fig}, right plot.

\WN

That's certainly fair. There is no direct way to tell what they did
for estimation. We trained estimation using the PSP. In that course
we taught estimation by parts.  Break the component into small parts,
use a different size tables for each part type, then combine the
results. 

In my experience, estimation by parts as taught in PSP was seldom
used in the field, it was too elaborate. Often they made tables using
the wbs level components. Not everyone estimated size (new and changed
lines of code) because the correlation of size and effort was
sometimes weak. For example, the effort to make a small change to a
large component didn't predict the required effort very well. For
that kind of problem, effort depended more on the complexity and
number of components. Other tasks, such as running build and test
depend more on defects found than changed code. So often they used
historical data to build tables of direct task effort rather than
make tables of the product size and run a regression. But in the end
there was nothing stopping them from estimating directly, and this
was probably the norm for non-development tasks.  You might see some
traces of using direct estimates in Figure~\ref{est_act:fig}. 

Looking at the left side of Figure~\ref{est_act:fig} you might notice
the huge variation of the actual from the estimate.  As long as you
mostly got things into the right bin and the high and low misses
balanced out the central limit theorem drives the overall estimate
toward the middle. That wide variation is really very typical in most
point estimates and I don't think it's been handled appropriately
most of the time. There are significant differences between people,
but people have a high variance too. If you don't account for both
the between variance effects are impassible to see. If you don't
account for the within variance you wonder why so many people don't
seem to follow a pattern.  If you don't account for both within and
between variation, your models will be fundamentally wrong.  

More practically, we were not concerned that every estimate was
right, only that the highs and lows would cancel out.  If we did this
correctly and used relevant estimation tables, we could avoid
systematic bias and the central limit theorem would be our friend.

\section{Data quality}

\DJ

People make mistakes during data entry, and their memory for past
events may be fuzzy.  How reliable and accurate are the values
contained in the CESAW dataset?

A previous analysis \cite{Shirai_14a} of the CESAW data investigated
the percentage of values thought unlikely to occur in practice (this
previous analysis included the data in the file {\itshape
CESAW\_defect\_facts.csv}).  The data quality analysis in this
section applies to the files: {\itshape CESAW\_task\_facts.csv} and
{\itshape CESAW\_time\_facts.csv}, i.e., the files analyzed.

There are a fixed number of minutes in a day, and people are unlikely
to be working during all of them.  While TSP practices recommend an
upper limit on the amount of continuous time spent on a task, there
may be deadlines that need to be met.

\begin{figure}
\begin{center}
\includegraphics{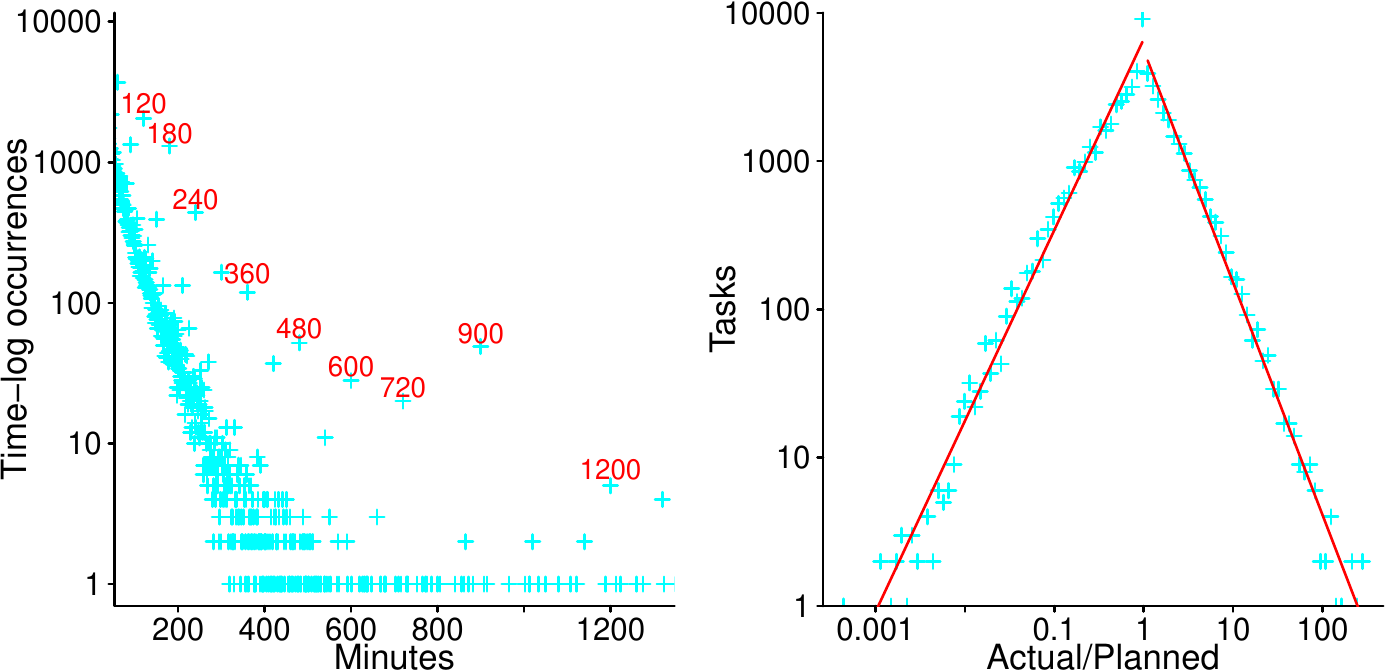}
\end{center}
\caption{Both plots are based on data from all projects.  Left: Number of tasks taking a given number of delta minute in the time log; right: Number of tasks taking a given actual to estimated ratio (ratios aggregated into 100 bins), red lines are power law fits with exponents 1.3 and -1.6.}
\label{delta_mins:fig}
\end{figure}

Figure~\ref{delta_mins:fig}, left plot, shows the number of tasks
recorded as taking a given number of delta minutes, in the time log,
over all projects.  The descending bundle of points crosses the
x-axis close to the number of minutes in a working day, i.e., 450.
The numbered points, in red, are round-numbers, and are suggestive of
people working for a planned amount of time.  The larger
round-numbers (e.g., 10, 12 and 15 hours) may be from continuous work
on one task, or forgetting to press the stop button after finishing a
task (and pressing it later just before going home).  The even larger
numbers may be due to pressing the stop button the next day.

Inaccurate data can cause analysis code to fail, in a statistical
sense.  For instance, the task facts file contains 638 (1\%) rows
where the start date is later than the end date.

The percentage of tasks that appear to be worked on continuously for
long periods is sufficiently small (0.27\% for 6-hours, 0.075\% for
10 hours), that they can be treated as background noise.

The process used to log the SiP dataset \cite{Jones_19a} checked that
each person's daily actual time equalled one working day.

\WN

As a coach, when I see unusually long hours, I try to verify them.
You always want to try to understand if something unusual is
happening, or if there was some sort of recording error. It's not
hard to image reasons for an unusually hard push, a severe bug or a
push for a key delivery. These should be rare.  As we get far removed
from the source, we have to take some care that our analysis is
robust to an occasional outlier.  Since software is written by
people, lots of odd or unexpected things can happen. 

\DJ

A large difference between the actual and estimated time for a task
may be due to one of the values being incorrectly recorded, it could
also be due to the implementation turning out to be much
easier/harder than expected, or a poor estimation.

Figure~\ref{delta_mins:fig}, right plot, shows the number of tasks
having a given actual to estimated ratio, for all projects (ratios
aggregated into 100 bins); red lines are power law fits with
exponents of 1.3 and -1.6.

Most task ratios are close to one, and the consistent pattern of
decreasing counts suggests that the more extreme values might not be
outliers.

Bill: Any ideas why some ratios are so extreme?
There are not that many, so perhaps the regular pattern is a regular
pattern of incorrect data entry?

\WN

I have a few guesses about the outliers, but I'd be sceptical of data
entry errors. Really big errors seem likely to be caught during
status reviews (unless the total is really small).  Very low values
are likely tasks that were stopped because they were de-prioritized
and removed from active work. Just closing them would not be my
choice because we tended to use the task close information for
progress tracking, but each team has their own way of working.  It's
surprising that some of the high ratios are as high as 100 to 1.  I'd
be interested in looking at task log comments to see what's
happening, for example some might be difficult to find bug. Sometimes
they used a task as a catchall for miscellaneous work, or they might
have put in an unrealistically small estimate for some reason.
Usually I'd expect replanning if a task went way beyond plan.  It's
remarkable that the shape slopes are so consistent on the log scale. 

\vfill

\section{General discussion}

\DJ

Companies want to control the processes they use, which is only
possible when they understand what is going on. Patterns of behavior
discovered by the analysis of historical data can help refine
existing understanding of development processes or suggest new ones.

This analysis has found many patterns in the CESAW data.  The
usefulness of this analysis of the CESAW data can only be assessed by
those involved in the production of software systems.  

\WN

There are two parts to this, the tracking and the post action
analysis. Both are really important.  The tracking helps steer
projects, make interventions, adjust priorities, control the costs,
hit targets. Post action analysis is how we get better for the next
time.

Just getting tasks to a small size and estimating with reasonable
accuracy is a huge win. The developer of the Process Dashboard
presented some evidence from Monte Carlo that just balancing work
frequently (as with Agile pull systems) was the single biggest source
of schedule improvement. It keeps everyone busy and identifies
problems early. 

Another big part of the data is understanding what really works and
what doesn't. CESAW showed that static analysis was cost-effective,
but really only about 1/3 as effective as a good review.  But it also
took only about 1/3 the time. 

\subsection{What did I learn?}

\DJ

While some of the patterns of behavior revealed by this analysis were
familiar to me, and I hope also others involved in software projects,
many were new to me.

\begin{itemize}
\item I was surprised by the short duration of many tasks.
If the recording processes is easy to use and management requires
detailed reporting, then here is the evidence that information on
short duration tasks can be captured,

\item this analysis is my first hands-on encounter with projects that
use the Work Breakdown Structure.  While it is possible to highlight
a variety of patterns that appear in this WBS data, without prior
experience of using WBS it is not possible for me to know whether
these patterns have any practical significance,

\item use of round-numbers, as estimation values, was so common that
model fitting sometimes had to treat them as a separate cluster of
values, e.g., see figure~\ref{interrupt_freq:fig}.

Estimates have to be sold to clients, and if the client thinks in
round-numbers a non-round number estimate introduces friction to the
sales process\cite{Loschelder_16}.

\end{itemize}

\WN

These are late estimates rather than early estimates used to bid a
job. I've found again and again that we have to stop the developers
from trying to estimate too precisely. They always want to use more
bins. But our data shows that it just makes things worse. If you keep
the components small, put it in the right bin most of the time
(accurate, not precise), then the central limit theorem will drive
the aggregate toward the middle.

\bibliography{SEMPR-main}

\end{document}